\documentclass[structabstract]{aa}  
\newcommand{\Ni}{$^{56}$Ni}
\newcommand{\Co}{$^{56}$Co}
\newcommand{\Fe}{$^{56}$Fe}
\newcommand{\arp}{Ar$^+$}
\newcommand{\hep}{He$^+$}
\newcommand{\nep}{Ne$^+$}

\newcommand{\Ms}{M$_{\odot}$}

\newcommand{\grays}{$\gamma$-rays}
\newcommand{\gray}{$\gamma$-ray}
\newcommand{\cmc}{cm$^{-3}$}
\newcommand{\kms}{kms$^{-1}$}

\newcommand{\fos}{Mg$_2$SiO$_4$}
\newcommand{\alu}{Al$_2$O$_3$}
\newcommand{\mic}{$\mu$m}
\usepackage{graphicx}

\usepackage{color}
\usepackage{txfonts}
\begin{document}

\title{Condensation of dust in the ejecta of type II-P supernovae} 
\author{Arkaprabha Sarangi \& Isabelle Cherchneff}
 \institute{Departement Physik, Universit{\"a}t Basel, Klingelbergstrasse 82, 4056 Basel, Switzerland }

   \date{Submitted 12 September 2014; Accepted 16 December 2014}
    \abstract
   {}
   {We study the production of dust in Type II-P supernova ejecta by coupling the gas-phase chemistry to the dust nucleation and condensation phases. We consider two supernova progenitor masses with homogeneous and clumpy ejecta to assess the chemical type and quantity of dust that forms. Grain size distributions are derived for all dust components as a function of post-explosion time.  }
   {The chemistry of the gas phase and the simultaneous formation of dust clusters are described by a chemical network that includes all possible processes that are efficient at high gas temperatures and densities. The formation of key bimolecular species (e.g., CO, SiO) and dust clusters of silicates, alumina, silica, metal carbides, metal sulphides, pure metals, and amorphous carbon is considered. A set of stiff, coupled, ordinary, differential equations is solved for the gas conditions pertaining to supernova explosions. These master equations are coupled to a dust condensation formalism based on Brownian coagulation.  }
   {We find that Type II-P supernovae produce dust grains of various chemical compositions and size distributions as a function of post-explosion time. The grain size distributions gain in complexity with time, are slewed towards large grains, and differ from the usual Mathis, Rumpl, \& Nordsieck power-law distribution characterising interstellar dust. Gas density enhancements in the form of ejecta clumps strongly affect the chemical composition of dust and the grain size distributions. Some dust type, such as forsterite and pure metallic grains, are highly dependent on clumpiness. Specifically, a clumpy ejecta produces large grains over 0.1 \mic, and the final dust mass for the 19 \Ms\ progenitor reaches 0.14 \Ms. Clumps also favour the formation of specific molecules, such as CO$_2$, in the oxygen-rich zones. Conversely, the carbon and alumina dust masses are primarily controlled by the mass yields of alumina and carbon in the ejecta zones where the dust is produced. The supernova progenitor mass and the \Ni\ mass also affect dust production. Our results highlight that dust synthesis in Type II-P supernovae is not a single and simple process, as often assumed. Several dust components form in the ejecta over time and the total dust mass gradually builds up over a time span around three to five years post-outburst. This gradual growth provides a possible explanation for the discrepancy between the small amounts of dust formed at early post-explosion times and the high dust masses derived from recent observations of supernova remnants. }
   {}

   
   \authorrunning{Sarangi \& Cherchneff}
   \titlerunning{Dust condensation in supernovae}

\maketitle

 \section{Introduction} 
 
 %------------------- Table 3 -----------------------
\begin{table*}
\caption{Parameters for the clumpy ejecta model with a 19~~\Ms\ stellar progenitor taken from Jerkstrand et al. (2011) and SC13.}
\label{tab3}
\centering
\begin{tabular}{l c c c c c c c c}
\hline \hline
Ejecta zones & Zone 1A & Zone 1B & Zone 2 & Zone 3 & Zone 4 & Zone 5 & Zone 6 & Total\\
\hline
Zone mass in~\Ms & 0.11 & 0.302 & 1.68 & 0.141 & 0.486 & 0.774 & 0.358 & 3.85 \\
Clump number & 44 & 118 & 654 & 55 & 189 & 301 & 139 & 1500 \\
$f_c$& 2.9(-2) & 4.1(-3) & 7.3(-2) & 2.0(-2) & 2.0(-2) & 1.5(-2) & 1.5(-2) & --\\
$n_c$(day 100) in \cmc & 2.47(12) & 2.83(13) & 2.11(12) & 8.60(12) & 1.26(13) & 4.24(13) & 4.29(13) & --\\
\hline
\end{tabular}
\tablefoot{Each clump has a mass of $2.6\times 10^{-3}$~\Ms.}
\end{table*}
%-----------------------------------------
 
 Important sources of cosmic dust include the explosion of supergiant stars as Type II supernovae (SNe). Dust formation was observed in SN1987A at infrared (IR) wavelengths a few hundred days after the explosion (\cite{luc89, dan91, wood93}), and this scenario has since been observed in several other Type II-P SNe (e.g., \cite{sug06, ko09, gal12, sza13}). Analysis of the IR flux emitted by this warm dust has indicated fairly modest amounts of solid condensates in the ejecta, in the $10^{-5}-10^{-3}$~\Ms\ range, while in the specific case of SN2003gd, a mass of 0.02~\Ms\ at day 678 after outburst was inferred from Spitzer data, under the assumption that the ejecta was clumpy. However, Meikle et al. (2007) contest this high dust mass 660 days post-explosion by explaining that an IR echo from pre-existing circumstellar dust could contribute to the emission.
 
Recent observations of SN remnants (Thereafter SNRs) in the sub-millimetre (submm) with AKARI, Herschel and ALMA have brought evidence of large reservoirs of cool ejecta dust, specifically in the 330-year-old SNR Cas~A, the 960-year-old Crab Nebula, and the young remnant of SN1987A. In Cas~A, the dust mass estimated from AKARI and Herschel data ranges from 0.06~\Ms\ (\cite{sib10}) to 0.075~\Ms\ (\cite{bar10}). In the Crab Nebula, $0.02$ to $0.2$ \Ms\ of dust were inferred from Herschel data (\cite{gom12, tem13}). For SN1987A, very high dust masses in the range $0.4-0.8$~\Ms\ were inferred from the submm fluxes measured with Herschel (Matsuura et al. 2012, 2014), and the mass was reduced to a lower limit of 0.2~\Ms\ from the analysis of new ALMA data (\cite{ind14}). These observations clearly indicate a much higher mass of dust than that detected a few years after the outburst. 

Whether these high dust masses are formed  in the ejecta but remained undetected, or the dust continues to grow to high mass values in the remnant phase decades after the explosion is a highly debated issue. It has recently been proposed that dust cluster growth occured over a time span of a few years after the SN explosion (\cite{sar13}, thereafter SC13). Furthermore, the growth of grains could not be ongoing after the nebular phase because the atomic and molecular accretion on the surface of dust grains could not proceed owing to a shortage of accreting species and the long time scale for accretion. These results contrast with a recent study by Wesson et al. (2015), who find the growth of dust grains continues from 1200 to 9200 days.  
 
Indirect evidence of the formation and growth of dust grains in SNe is provided by the study of pre-solar grains from meteorites. Some of those bear the isotopic anomaly signatures characteristic of SNe, and include the presence of radiogenic $^{44}$ Ca, which stems from the decay of short-lived $^{44}$Ti, an isotope only produced in SNe  (\cite{zin07}). Pre-solar grains of oxides, silicates, carbon, silicon carbide, silicon nitride, and silica formed in SN ejecta have been identified (\cite{zin07, hop10, haen13}). Typical lower limits for grain sizes are in the $0.1-1$~\mic\ range, with some evidence of very large grains; e.g., one SiC grain with a radius of 35 \mic\ has been identified (\cite{zin10}). The isotopic anomaly signatures of the pre-solar SN grains imply mixing in the ejecta, whereby the innermost and outermost zones might have been in contact during  or after the explosion. These results indicate the dust formed in the SN ejecta can survive the SNR phase, be incorporated to the Interstellar Medium, and travel to the solar system. 

Finally, the high masses of dust inferred from the observations of damped Ly$\alpha$ systems and quasars at high redshift (\cite{pei91,pet94}) hint at a possible contribution of massive SNe, because massive stars evolve on short time scales that are compatible with the age of the Universe at high redshift (\cite{dwek11}). Massive SNe do form high masses of dust grains in their ejecta (\cite{noz03, sch04, cher10}), but these grains are heavily reprocessed in subsequent evolutionary phases, for example, in the remnant phase and the Interstellar Medium. When interstellar dust destruction is considered and a top-heavy initial mass function is assumed, a dust mass of $\sim 1$ \Ms\ produced per SN is necessary to explain the high redshift dust (\cite{dwek07}). This value is higher than the dust masses derived from IR data of SNe and comparable, although higher, to the dust masses inferred from submm data of SNRs. It is thus paramount to shed light on the processes underpinning dust production in SNe, locally, and at high redshift. 

Following our study on the molecule and cluster formation in Type II-P SNe of various progenitor masses (SC13), we present an exhaustive model of the dust production in SNe ejecta, which includes the coupling of the gas-phase and nucleation phase chemistry to the condensation of dust grains up to $\sim 5 $ years after the SN explosion. Size distributions for the various dust components are derived, and we explore the effect of a low $^{56}$Ni mass and ejecta clumpiness on the synthesis of molecules and dust grains. The derivation of grain chemical compositions, masses, and size distributions is important to model the dust fluxes in the IR and submm, and to assess the chances of survival of SN grains in the SN remnant phase. In \S\ \ref{sec2}, we present the physical models used in the study, and in \S\ \ref{sec3}, the chemical processes and the formalism underpinning the nucleation and the condensation phase. The results for homogeneous and clumpy ejecta are presented in \S\ \ref{sec4}, and a discussion follows in \S\ \ref{sec5}. 

\section{Physical model of the supernova ejecta}
\label{sec2}

When a massive star explodes as a supernova, its helium core is crossed by the explosion blast wave that deposits energy to the gas. The reverse shock created at the base of the progenitor envelope propagates inward and triggers Rayleigh-Taylor instabilities and macroscopic mixing over time scales of a few days (\cite{jog10}). These instabilities result in the fragmentation of the He-core, but a chemical stratification persists over time. Radioactivity plays a crucial role in generating the light curve and impinges on the ejecta chemistry. The radioactive \Ni~produced in the explosion decays into \Co~on a time scale of a few days. In turn, \Co~decays into \Fe~with a half-life of $\sim$ 113 days. This decay sequence creates a flux of \gray~photons that pervades the ejecta. The degrading of \grays~to X-rays and ultraviolet (UV) photons occurs by Compton scattering and creates a population of fast Compton electrons in the ejecta. These fast electrons ionise the gas, and produce ions such as \arp, \nep, and \hep. These ions are detrimental to the survival of molecules in the ejecta gas (\cite{lepp90, cher09}, SC13). 

We consider homogeneous, stratified ejecta, whose elemental compositions are given for a 15 \Ms\ and 19~\Ms\ stellar progenitors (\cite{rau02}). The ejecta consists of mass zones of specific chemical compositions, which are summarised in Table \ref{tab1} of the Appendix A as a function of ejecta zoning. Each zone is microscopically mixed, and we assume no chemical leakage between the various zones. The ejecta gas temperature and density are derived from explosion models and vary with post-explosion time according to equations 1 and 2 in SC13 for both the 15 \Ms\ and 19~\Ms\ progenitors. For the gas density, a constant value of $1.1\times 10^{-11}$ g cm$^{-3}$ at day 100 is assumed in all He-core mass zones, hence the ejecta is homogeneous. Because each mass zone has a specific chemical composition, this constant gas density translates into different gas number densities at day 100 in the mass zones. The gas temperatures and number densities are listed in Table \ref{tab2} of the Appendix A for both progenitor masses. 

We also consider a non-homogeneous, clumpy ejecta for the 19~\Ms\ stellar progenitor, which we choose as a surrogate to SN1987A and other massive SNe. We build up a simple model for a clumpy ejecta as follows: for each zone, we use the volume filling factor $f_c$ derived by Jerkstrand et al. (2011) in their modelling of the ultraviolet, optical, and near-IR emission lines observed in SN1987A. We assume a fiducial number of 1500 for the total number of clumps in the ejecta, in agreement with radiative transfer models of the IR spectral energy distribution of various SN ejecta (e.g., \cite{gal12}). Assuming the He-core mass for the 19~\Ms\ stellar progenitor listed in Table \ref{tab2}, we derive a typical clump mass of $ \sim 2.6\times 10^{-3}$~\Ms\ for all clumps, a value which agrees well with typical clump masses derived from 3-D explosion models (e.g., \cite{ham10}). The enhancement over the homogeneous gas density at day 100 is then given by $1/f_c$, and the gas number density $n_c$ in the clumps is estimated as $n_c=n_0 / f_c$, where $n_o$ is the gas number density for the homogeneous zone. All parameters are listed in Table \ref{tab3} as a function of ejecta zones. The initial atomic yields are those of the 19~\Ms\ homogeneous model given in Table \ref{tab1}, and the gas number density follows a time variation as in SC13. We will see in \S~\ref{clump} that high amounts of CO form at early post-explosion time in the clumpy ejecta and will probably affect the cooling and temperature of the ejecta zones, but we do not consider molecular cooling at this stage of the modelling. Therefore, the gas temperature for the 19 \Ms\ clumpy case is that given in Table \ref{tab2}.

\section{Dust nucleation and condensation model}
\label{sec3}

%------------ fig 0 ----------------------
\begin{figure*}
   \includegraphics[width=\columnwidth]{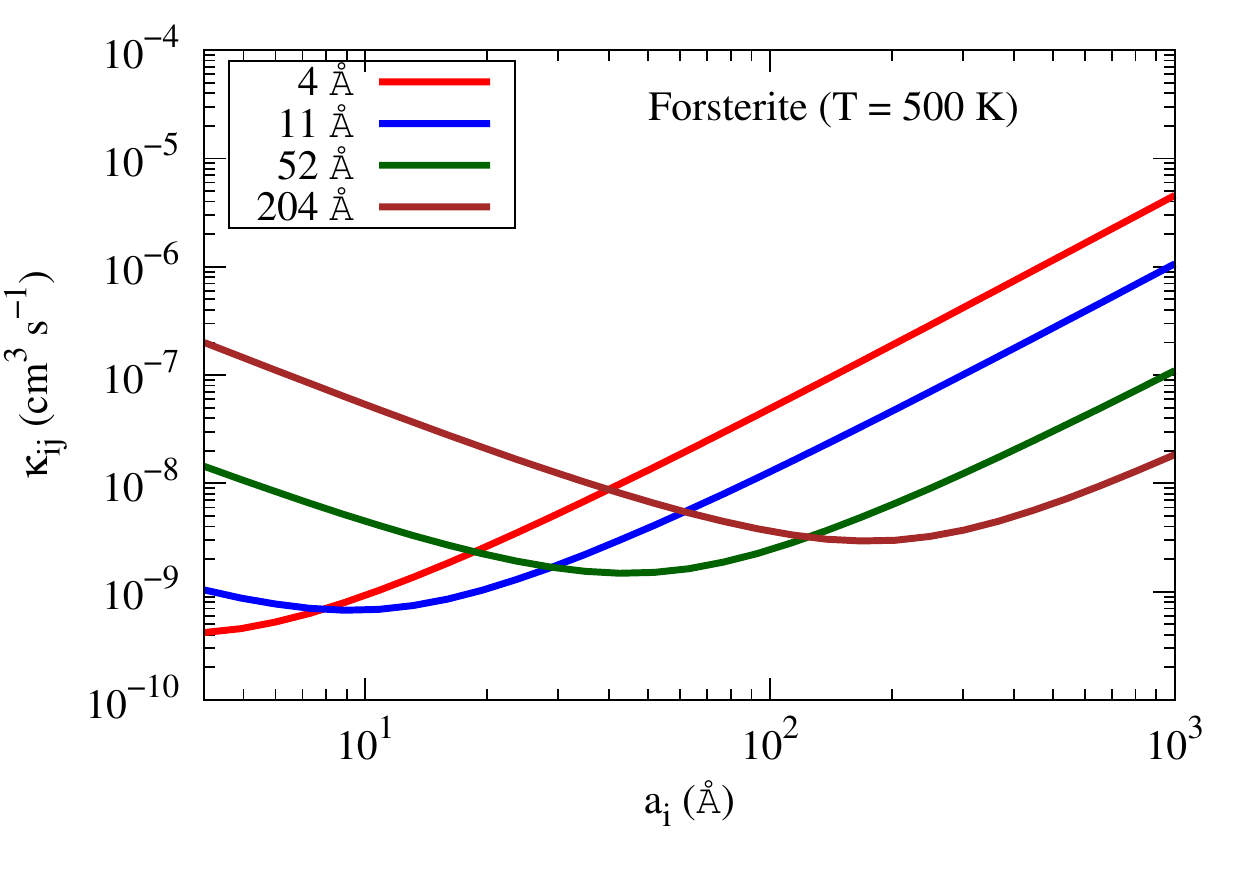}
    \includegraphics[width=\columnwidth]{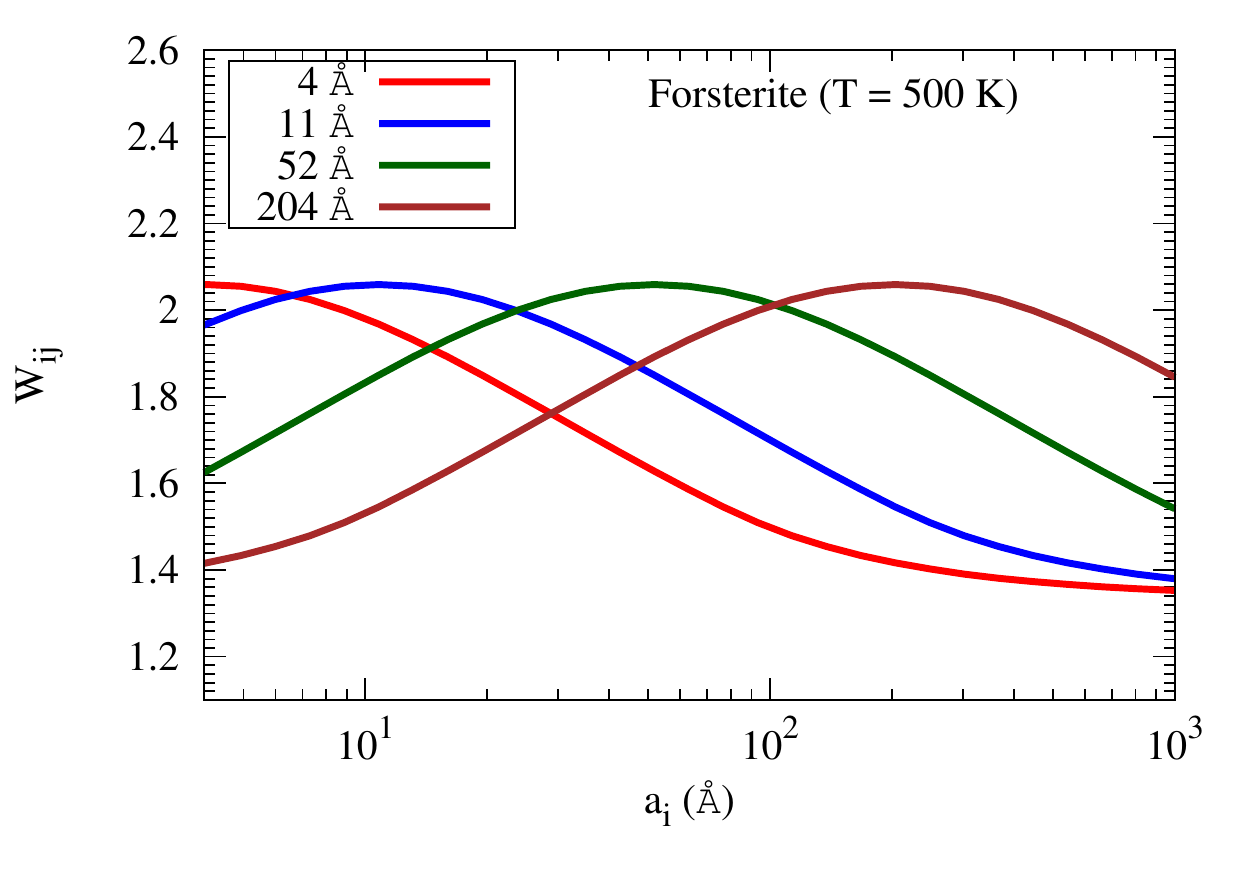} 
    \includegraphics[width=\columnwidth]{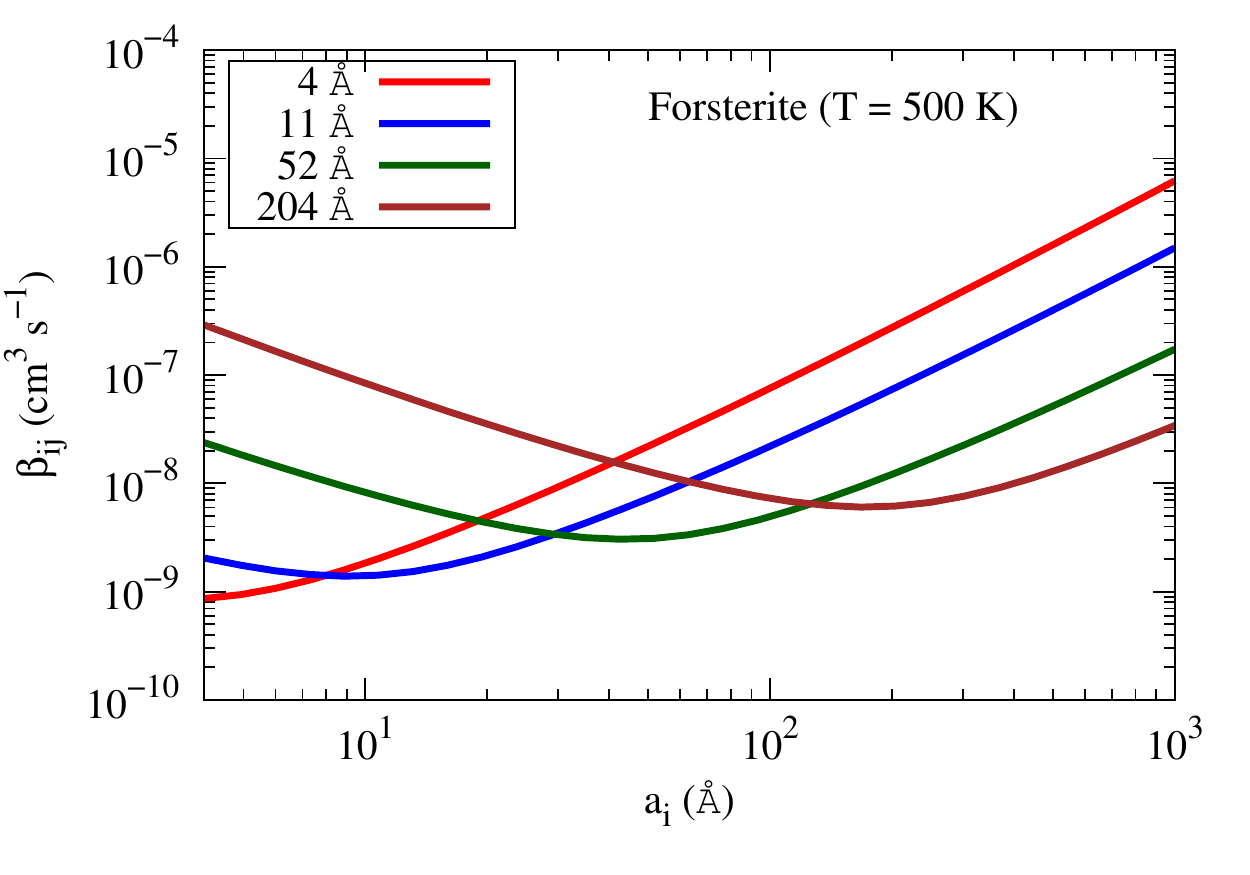} 
    \includegraphics[width=\columnwidth]{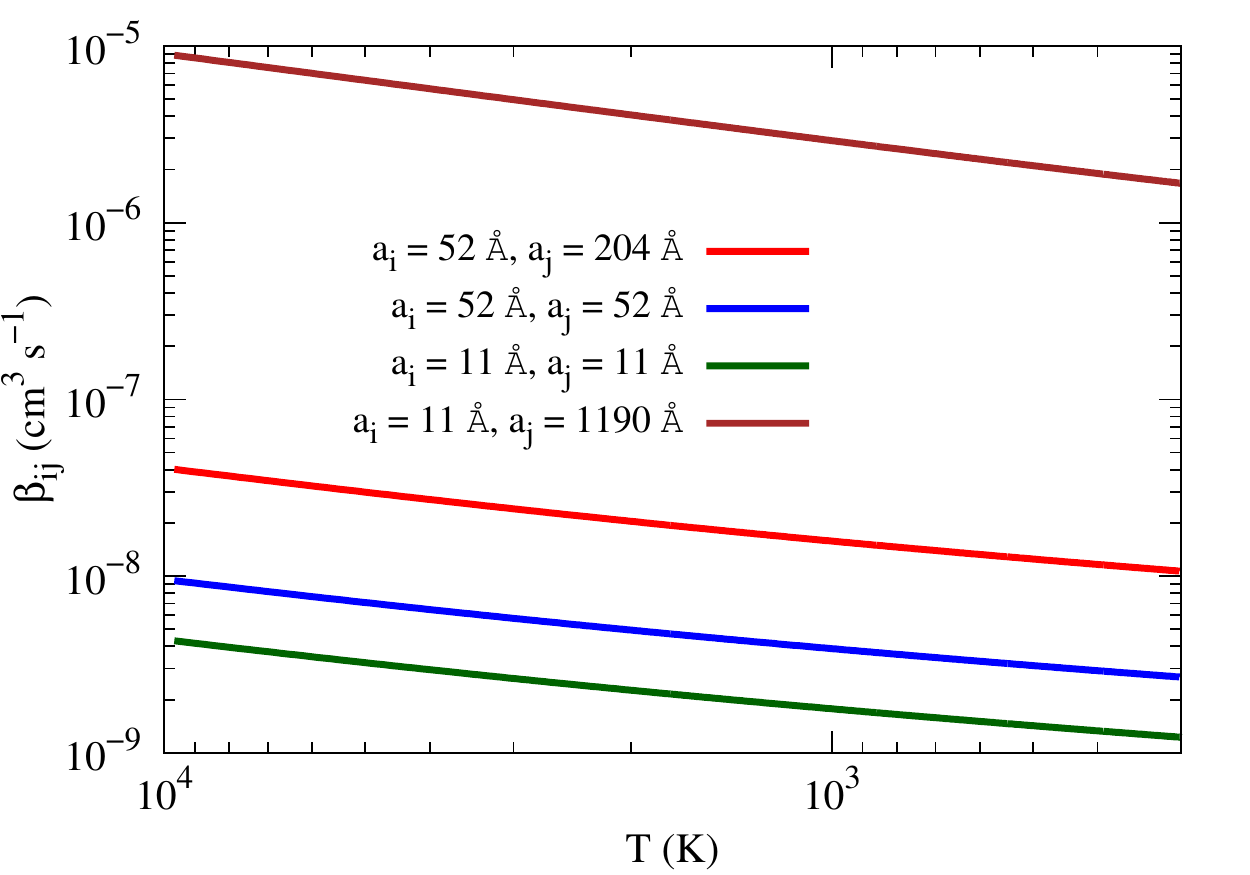} 
 \caption{Rate of Brownian diffusion $K_{ij}$ (top-left), coalescence enhancement factor $W_{ij}$ (top-right), Brownian coagulation rate $\beta_{ij} $(bottom-left) as a function of the collider radius $a_i$, for forsterite grains and four different values of the target size $a_j$. The gas temperature is $T_{gas} = 500 $K. The Brownian coagulation rate as a function of gas temperature in the ejecta (bottom-right) is shown from 100 to 1300 days for four sets of collider and target sizes given by the ($a_i$;$a_j$) parameters.}
 \label{fig0}
\end{figure*}
%--------------------------------------

The nucleation of small dust clusters out of the gas phase in the ejecta of Type II-P SNe was studied by SC13. The formation of dimers of forsterite (\fos) carbon, silicon carbide (SiC), pure metals that include silicon, iron, and magnesium, magnesia (MgO), iron sulphide (FeS), and iron oxide (FeO), has been modelled by using a chemical kinetic approach applied to the ejecta of SNe associated with stellar progenitors of various masses. In SC13, the formation of the molecule AlO was studied as an indicator of alumina formation, but no nucleation scheme for alumina clusters was implemented. As for carbon, the small carbon clusters were represented by the chain C$_{10}$. 

In this paper, the growth of alumina, \alu, is described by the dimerisation of AlO, followed by the oxidation of the dimer through reactions with oxygen-bearing molecules (O$_2$, SO), which leads to the formation of the small cluster \alu\ (\cite{bis14}). We then consider the tetramer of \alu, the (\alu)$_4$ molecule, to be the stable gas phase precursor of alumina dust. The growth of amorphous carbon is extended to carbon chains larger than C$_{10}$ and involves the synthesis of carbon chains through reactions with atomic carbon and C$_2$, the closure of these chains as rings when the number of carbon atoms N$_C \ge 10$, and the growth of rings, which is controlled by C$_2$ addition. The first stable cage structure that subsequently forms is C$_{28}$, and the cage keeps growing through C$_2$ addition to the carbon lattice without cage fragmentation (\cite{dun12}). The C$_2$ inclusion can proceed towards forming the fullerene cage, C$_{60}$, but the efficiency at synthesising fullerene depends on the gas number densities and the abundance of the growing agent, C$_2$. As in SC13, we consider the synthesis of pure metallic grains by forming the small tetramers (SiC)$_4$, (Si)$_4$, (Fe)$_4$, (FeS)$_4$, and the formation of the pentamer (SiO)$_5$ as a small cluster precursor of silica. However, the nucleation and condensation of silica is not considered in the present study because of the paucity of information available on these processes. 

These molecular clusters act as dust seeds in the condensation phase, where they grow by coagulation to form large grains. Growth through accretion of atoms and molecules on the surface of dust grains may also occur if there is enough accreting species in the gas phase (i.e., Si, SiO, C, or C$_2$ depending on the ejecta zone) and available grain surfaces to accrete on. By the time small dust clusters have formed from the gas phase, the supply of growing gas-phase species has severely decreased in the dust-forming ejecta zones, as shown in SC13. Accretion can thus not proceed because of this shortage of accreting species, and we consider coagulation of small dust clusters the only dust growth mechanism before day 2000.

To describe dust coagulation, we use the formalism developed by Jacobson (2005) (thereafter J05), where the variation with time of the number density of a grain of specific volume $v$ is described by the integro-differential coagulation equation given by 

%----------------- eq1 ---------------------
\begin{equation} 
\label{eq1}
\frac{dn_v (t)}{dt} = \frac{1}{2} \int\limits_{v_0}^v \beta_{v-v',v} n_{v-v'}n_{v'} dv' - n_{v} \int\limits_{v_0}^\infty \beta_{v,v'} n_{v'} dv'.
\end{equation}
% ---------------- eq1 ----------------------
Here, $t$ is the time, $v'$ and $(v-v')$ are the volumes of the two coagulating particles, n$_v$ is the number density of grains with volume $v$, $\beta_{v, v'}'$ is the rate coefficient of coagulation between particles with volume $v$ and $v'$, and $v_0$ is the volume of the largest gas-phase cluster seed. The rate of coagulation $\beta_{v_i,v_j}$ for particles $i$ and $j$ is controlled by physical processes such as Brownian diffusion, convective Brownian motion enhancement, gravitational collection, turbulent inertial motion, and Van der Waal's forces. Typical gas number densities in SN ejecta at 300 days post-explosion range between $10^9-10^{11}$ \cmc. Therefore, the ejecta gas is characterised by a free-molecular regime, which is defined by $\lambda_p \gg a_i $, where $\lambda_p$ is the mean free-path of a particle in the gas and $a_i$ is the radius of the $i^{th}$particle. In a free-molecular regime, Brownian diffusion prevails in the coagulation process, whereas the other processes are relevant in case of larger particles and denser media. Brownian diffusion accounts for the scattering, collision, and coalescence of the grains through Brownian motion. 

For the sake of simplicity, we rename the rate coefficient of coagulation $\beta_{v_i,v_j}$ as  $\beta_{ij}$ between two grains $i$ and $j$ of radius $a_i$ and $a_j$, respectively. The rate is given by 
%----------- eq2 ---------------------------
\begin{equation} 
\label{eq2}
\beta_{ij} = \frac{4\pi(a_i + a_j)(D_{c,i} + D_{c,j})W_{i,j}}{\frac{(a_i + a_j)}{(a_i + a_j) + \sqrt{\delta_i^2 + \delta_j^2}} + \frac{4(D_{c,i} + D_{c,j})}{(a_i + a_j)\sqrt{v_{p,i}^2 + v_{p,j}^2}}}
\end{equation}
% ----------- eq2 -------------------------

where $D_{c,i}$ \& $D_{c,j}$ are the diffusion coefficients for particle $i$ and $j$, respectively, $v_{p,m}$ is the mean thermal velocity for particle $m$, $\delta_i$ is the mean distance of particle $i$ from the centre of a sphere traveling a distance $\lambda_p$, and $W_{ij}$ is the enhancement factor due to the effect of Van der Waal's dispersion forces (J05, \cite{sp06}). The formalism assumes the dust grains have a temperature similar to that of the local gas of the ejecta zone where the grains form at a specific epoch. In the free-molecular regime, we have $\sqrt{\delta_i^2 + \delta_j^2} \gg (a_i + a_j)$, and Equation \ref{eq2} reduces to 
% ---------eq3 --------------------------
\begin{equation} 
\label{eq3}
\beta_{ij} = K_{ij} \times W_{ij} = \pi(a_i + a_j)^2 \sqrt{v_{p,i}^2 + v_{p,j}^2} W_{ij},
\end{equation}
% ----------- eq3-------------------------

where $K_{ij}$ is the Brownian diffusion term. The Van der Waal's forces develop weak, local charge fluctuations and enhance the rate of coagulation for particles with size in the molecular range, leading to the enhancement factor $W_{ij}$. The interaction potential $V(r)$ between two particles separated by a distance $r$ is defined by the Hamaker's theory and using London dispersion forces (\cite{st89, al87}), and is given by 

%------------- eq4 ----------------------
\begin{eqnarray}
\label{eq4}
V(r)& = & - {kT}\times \frac{A'}{12} \bigg(\frac{1}{(\frac{r}{a_i + a_j})^2-1} + \frac{1}{(\frac{r}{a_i + a_j})^2} \\ \nonumber
& & + 2\ln \bigg(1- \big(\frac{a_i + a_j}{r}\big)^{2}  \bigg)\bigg) \\
\end{eqnarray}
%------------------------------------
where $T$ is the gas temperature, and $A'$ is given by 
%---------------eq 5 ------------------------
\begin{equation} 
\label{eq5}
A' = \frac{A}{kT} \frac{4a_ia_j}{(a_i + a_j)^2}.
\end{equation}
%-----------------------------------------
In the above equation, A is the Hamaker constant which varies according to the physical properties of individual dust species. The enhancement factor $W_{i,j}$ due to the Van der Waal's dispersion forces is given by
%---------------- eq6 -------------------------
\begin{equation}
\label{eq6}
W_{ij} = \bigg(\frac{r_T}{a_i + a_j}\bigg)^2 e^{\frac{-V_{ij}(r_T)}{kT}}
\end{equation}
%-----------------------------------------------
where $r_T $ is the minimum separation between the two particles when the difference between attractive and repulsive potentials reaches a minimum, described by $\frac{\partial}{\partial r}(V(r)-2kT\ln r) = 0$ (\cite{st89}).

According to Equation \ref{eq3}, the coagulation rate, $\beta_{ij}$, is defined as the product of the Brownian diffusion term $K_{ij}$ and the enhancement factor due to coalescence $W_{ij}$. The Brownian diffusion term is controlled by the grain temperature, assumed to be equal to the gas temperature, while $W_{i,j}$ depends on the gas temperature. Both terms also depend on the ratio of the collider radii $f_a = a_i / a_j$, as shown in Figure \ref{fig0}, where the dependence of $K_{ij}$, $W_{ij}$, and $\beta_{ij}$ on collider and target sizes are shown. The diffusion term is large when $f_a \gg 1$ and reaches a minimum when $f_a \simeq1$. Conversely, the value taken by $W_{ij}$ is high for $f_a \rightarrow 1$, but is confined to a small range for all values of ($a_i$,$a_j$). Therefore, the rate of Brownian coagulation $\beta_{ij}$ is controlled by the variation of $K_{ij}$, as seen from Figure \ref{fig0}. The coagulation rate is minimal when the two colliders are of the same size, and is high when $f_a$ is large. Therefore, the large grains present in the gas efficiently coagulate with the newly formed small grains, and this efficiency warrants the growth of large dust grains in the ejecta. 

The evolution of the coagulation rate $\beta_{ij}$ is shown as a function of gas temperature in the ejecta and for specific pairs of collider sizes in Figure \ref{fig0} (bottom-right panel). Coagulation becomes less efficient as the gas temperature decreases, and typical rate values for $\beta_{ij}$ are in the range $10^{-9}-10^{-4}$ \cmc~s$^{-1}$ for colliders and targets of all sizes and ejecta gas temperatures between 100 and 2000 days post-explosion. 

%rate of Brownian coagulation $\beta_{ij}$ is directly proportional to $f_a$, i.e., the process is more efficient when $a_i\gg a_j$, as shown in Figure \ref{fig0}, where both the coagulation rate (top panel) and the coalescence factor (bottom panel) are shown as a function of $a_i$ and $a_j$. Conversely, the coalescence factor $W_{ij}$ is inversely proportional to $f_a$ and reaches a maximum when $f_a \rightarrow 1$. $W_{ij}$ acts as an enhancement factor to $\beta_{ij}$, with a value comprised between 1 and 5 depending on the gas temperature and the Hamaker constant. However the rate coefficient is dominated by Brownian coagulation. The coagulation rate $\beta_{ij}$ is thus minimum when $a_i = a_j$, i.e., 
A semi-implicit volume conserved model has been developed to solve Equation \ref{eq1} (J05, \cite{sp06}). Dust grains are assumed to maintain a spherical morphology and compact structure. The grains of individual dust species are assigned to discrete bins, following a volume ratio distribution given by $v_n = \gamma^{n-1} v_0$, where $v_n$ is the volume of the $n^{th}$ bin, $\gamma$ is a constant defined as the ratio of the volumes of adjacent bins, and $v_0$ is the volume of the first bin, determined by the size of the gas-phase precursor. At each time step, the volume of the first bin, $v_0$ corresponds to the radius $a_0$ of the stable, largest clusters produced from chemical kinetics in the nucleation phase. Particles with volumes intermediate to any two consecutive bins are allocated following a volume fractionation formalism (\cite{mj94}). The various quantities are then calculated according to Equations \ref{eq3} $-$ \ref{eq6}, Equation \ref{eq1} is integrated, and the grain sizes of individual dust components are derived for each time step. Values for $a_0$ and the Hamaker coefficient $A$ are summarised in Table \ref{tab4} for the dust types considered in this study. 

%---------------------- fig1 -----------------------
\begin{figure}
   \resizebox{\hsize}{!}{\includegraphics{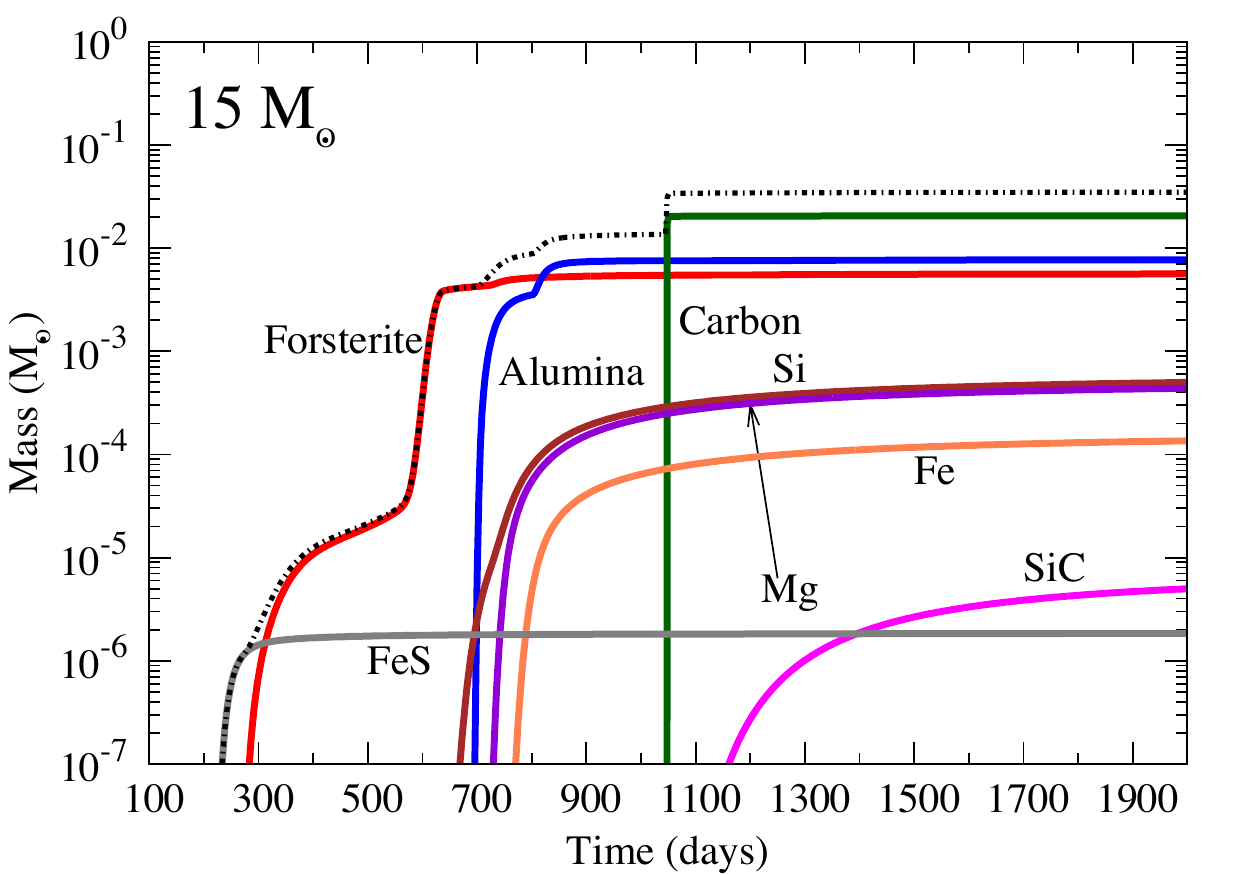}}
\caption{Dust mass for grains with radius larger than 10~\AA\ summed over all ejecta zones as a function of post-explosion time and dust type. The dotted grey line corresponds to the total dust mass formed in the homogeneous ejecta with progenitor mass 15~\Ms. }
\label{fig1}
\end{figure}

%-----------------------------------------------------

%--------------- Tab 4 -----------------------------------
\begin{table}
\caption{Initial grain size $a_0$ equivalent to the size of the largest dust cluster formed in the gas phase from chemical kinetics and the Hamaker constant $A$ for different dust components. }
\label{tab4} 
\centering
\begin{tabular}{l l l}
\hline \hline
Dust type &  $a_0$ (~\AA)& $A$ (10$^{-20}$ J) \\
\hline
Forsterite & 3.33 & 6.5\tablefootmark{a, b}  \\
Alumina & 3.45 & 15\tablefootmark{a, b, d}  \\
Carbon & 3.92 & 47 \tablefootmark{c}\\
Pure Magnesium & 2.29 & 30 \tablefootmark{c, d}\\
Silicon Carbide & 2.15 & 44 \tablefootmark{d} \\
Pure Silicon & 2.46 & 21 \tablefootmark{e} \\
Pure Iron & 2.81 & 30 \tablefootmark{c, d} \\
Iron Sulphide & 3.0 & 15 \tablefootmark{f} \\
\hline
\end{tabular}
%\tablefoot{
%\tablefoottext{a}{\cite{is92, ho07}}
\tablefoottext{a}{\cite{rn08},}\tablefoottext{b}{\cite{mm00},}\\
\tablefoottext{c}{\cite{ho07},}\tablefoottext{d}{\cite{is91},}\\
\tablefoottext{e}{\cite{fr95},}\tablefoottext{f}{\cite{bg97}.}
%}
\end{table}
%------------------------------------------
\section{Results}

%----------------------- figure 2 ----------------------------------

\begin{figure*}
 \centering
 \includegraphics[width=\columnwidth]{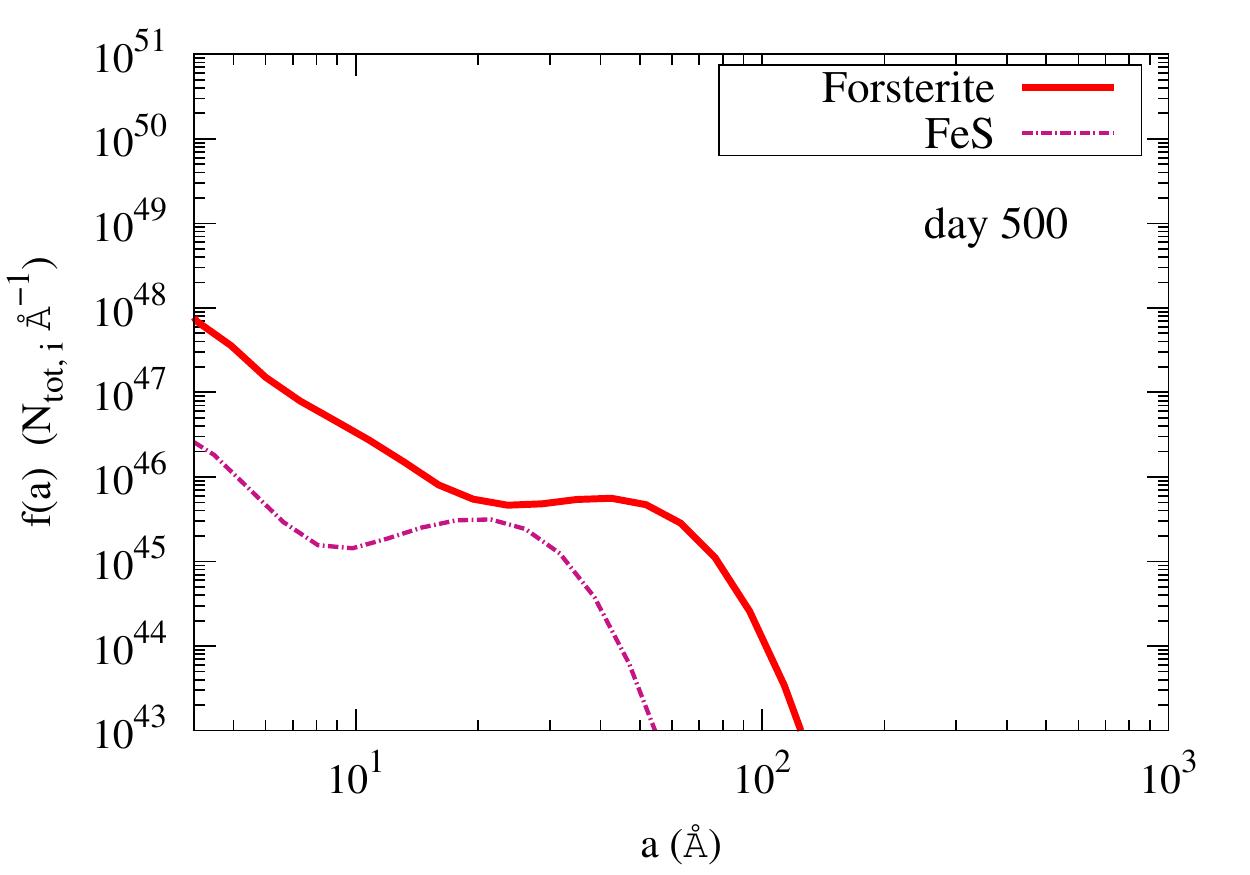} 
  \includegraphics[width=\columnwidth]{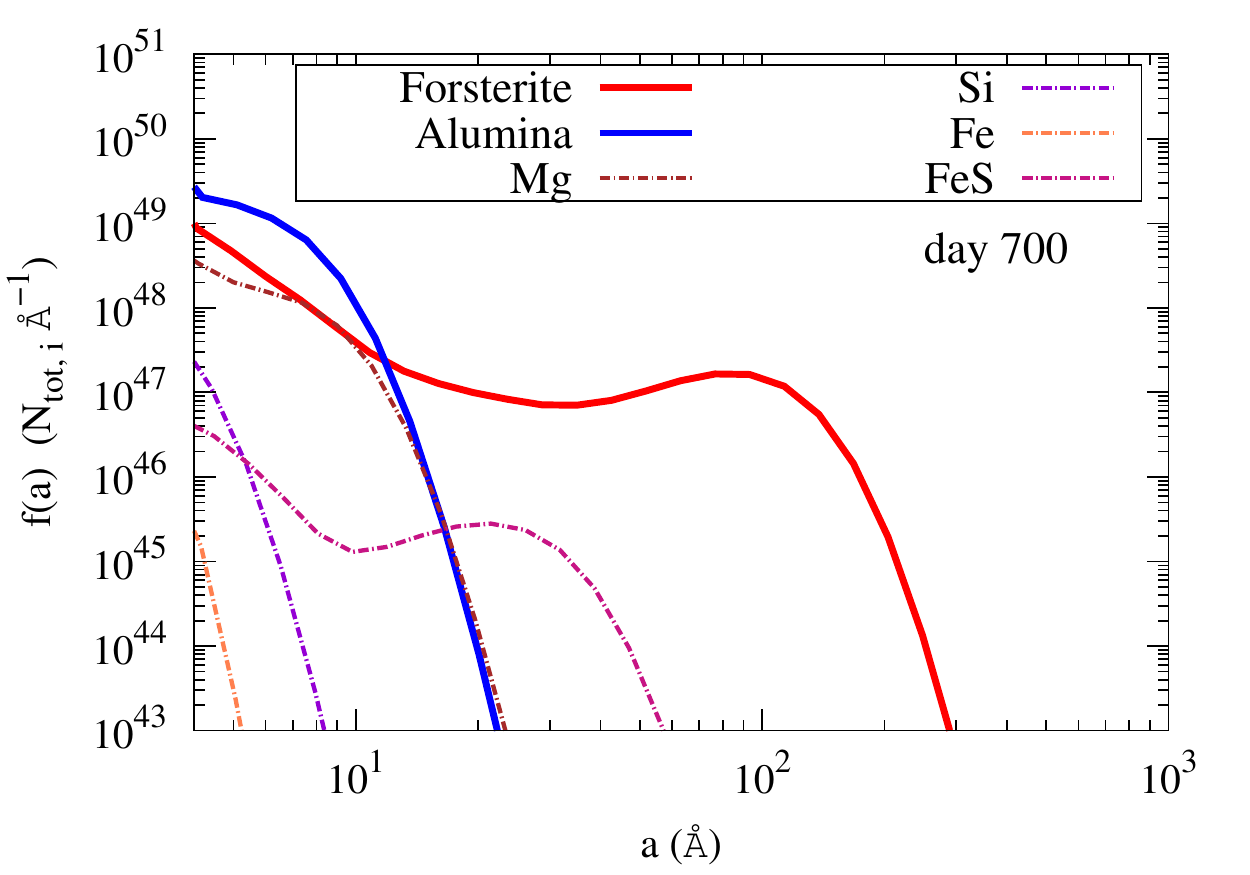} 
  \includegraphics[width=\columnwidth]{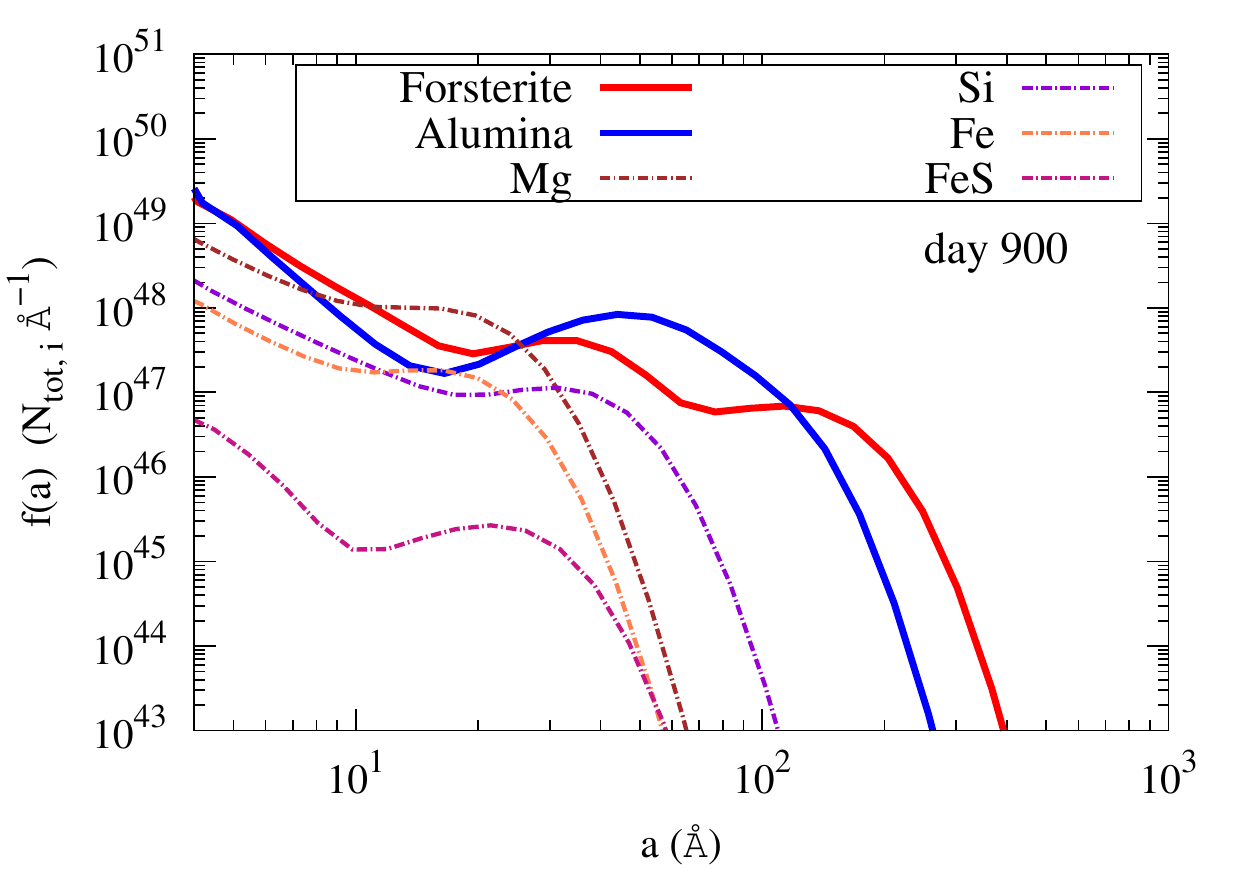} 
  \includegraphics[width=\columnwidth]{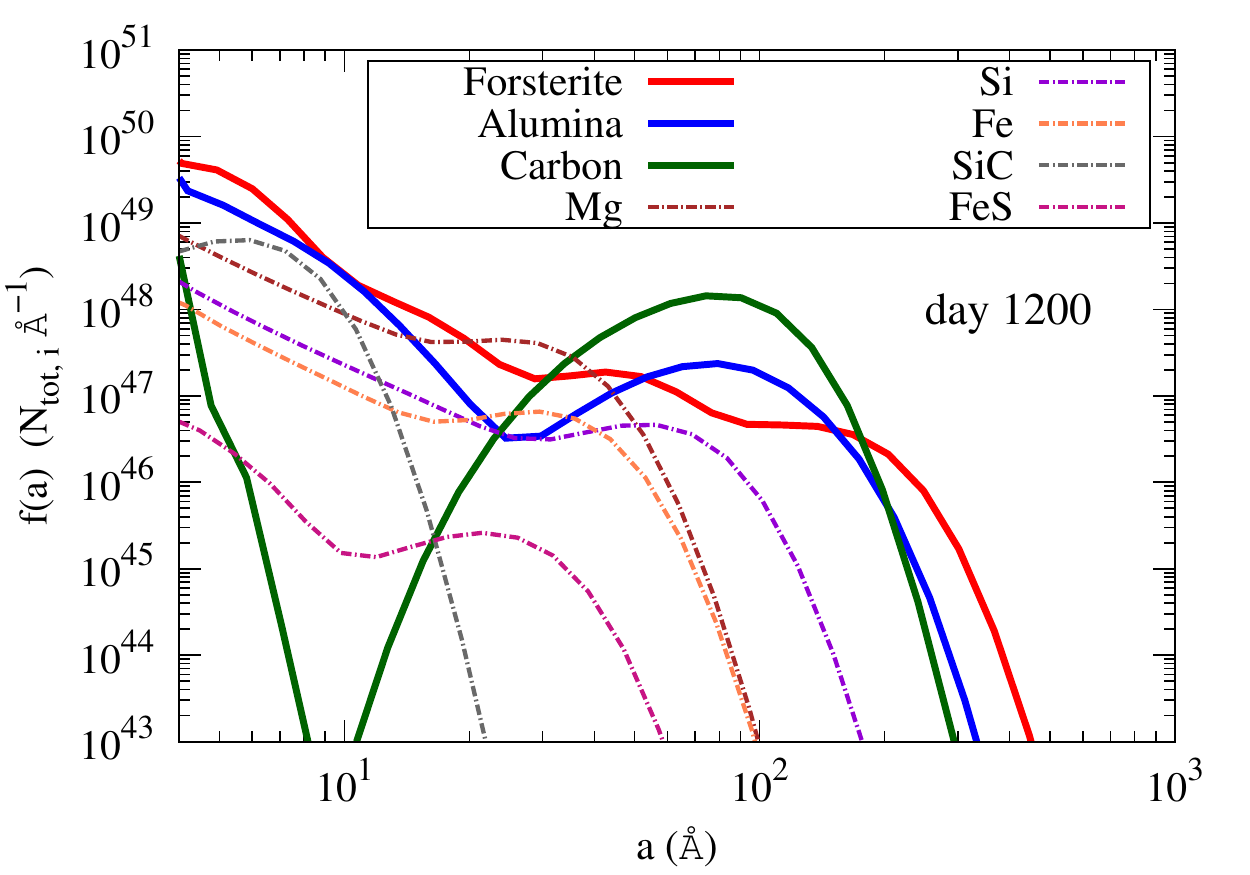}
  \includegraphics[width=\columnwidth]{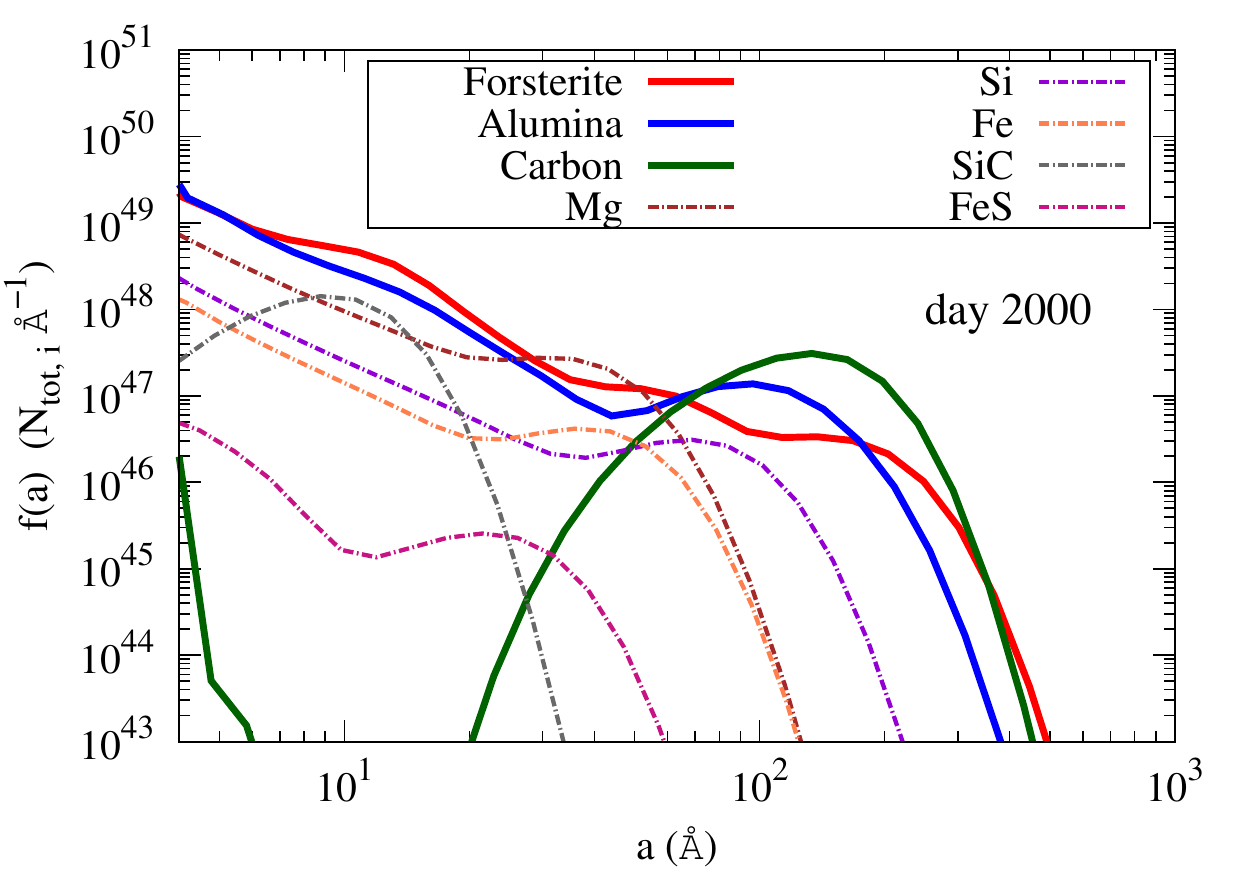}
   \includegraphics[width=\columnwidth]{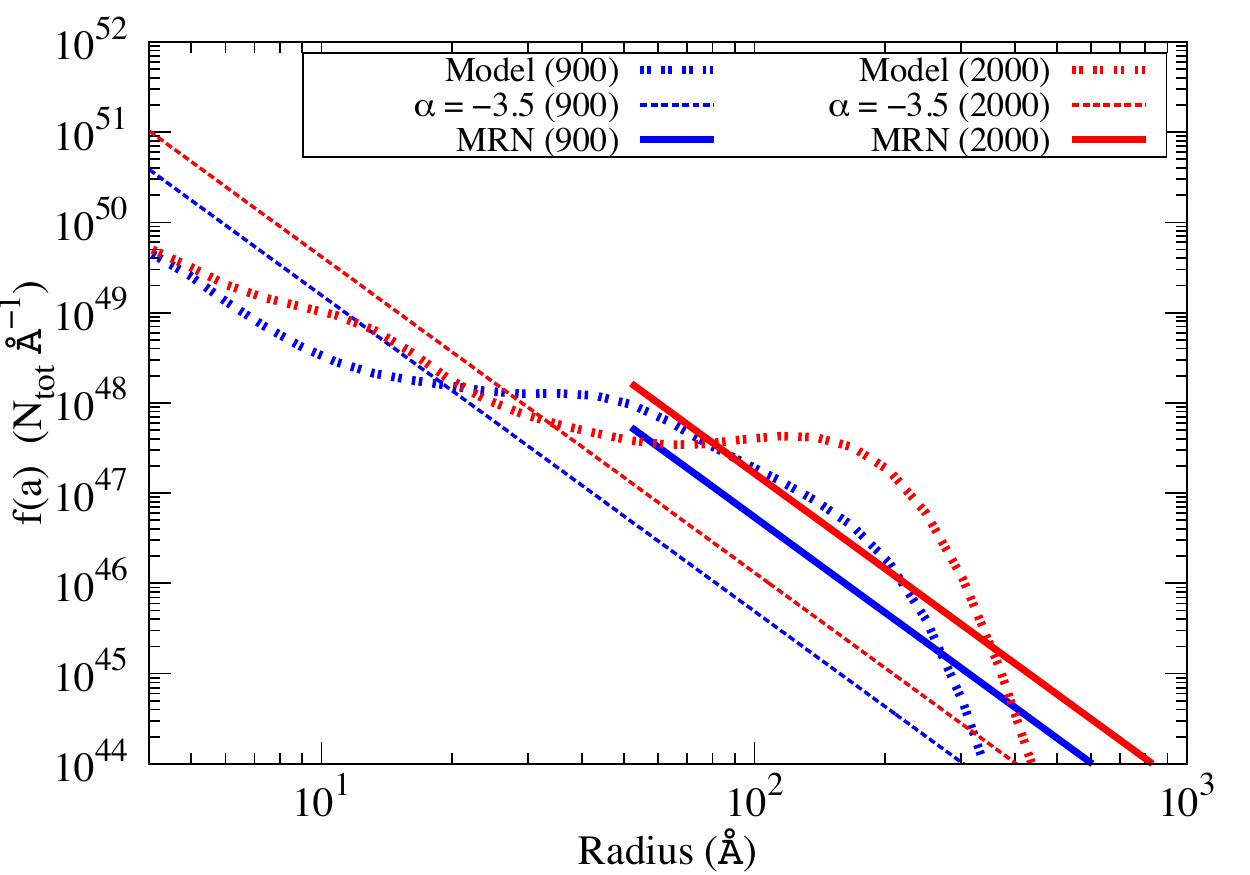}    
   \caption{Grain size distributions $f_i(a)$ as a function of post-explosion time for the various dust types formed in the homogeneous ejecta of a 15~\Ms\ SN. The bottom-right panel shows the grain size distribution summed over all dust types, a power law distribution with exponent -3.5 and the MRN power-law distribution (exponent -3.5 and $50$~\AA $ < a < 2500$ \AA) for an equivalent dust mass at day 900 and 2000 post-explosion. }
 \label{fig2}
\end{figure*}

 %------------------------------------------------------------------

\label{sec4}
The condensation phase is coupled to the nucleation phase to provide an exhaustive description of the synthesis of dust in Type II-P SN ejecta. We follow the gas phase chemistry by describing the formation of species (atoms, ions and molecules) and dust molecular clusters with a network of chemical reactions, as in SC13, and couple the chemistry to the simultaneous condensation of molecular clusters in dust grains from 100 to 2000 days after the SN explosion. Results on grain masses and size distributions for the various dust components are presented for a 15~\Ms\ stellar progenitor, characterised by a homogeneous ejecta and a $^{56}$Ni mass equals to $0.075$~\Ms\ (i.e., standard case). We further explore the effect of a low $^{56}$Ni mass (0.01~\Ms) on the condensation of dust grains.. We also consider a 19 \Ms\ stellar progenitor case as a surrogate of SN1987A and other massive SNe, for which we model the formation of dust in a homogeneous and a clumpy ejecta over a similar time span. The normalised size distributions for the 15 \Ms\ standard case and the 19 \Ms\ clumpy case are listed in Tables \ref{taba3}$-$\ref{taba5} in the Appendix A.

\subsection{15~\Ms\ progenitor: Standard Case} 
\label{sec1}

The SN ejecta of our standard case follows the stratification given in Table \ref{tab1}, where the zones are as in SC13. The Si/S zone (zone 1A) is conducive to the formation of iron sulphide, pure silicon and iron clusters. The oxygen core of the ejecta includes zones 1B, 2 and 3. Oxygen-rich dust components such as silicates and alumina are synthesised in this region. An excess of magnesium in the ejecta also leads to the formation of pure magnesium clusters. Zones 4A and 4B, where most of CO molecules form, have little contribution to the total dust mass. Finally, the outermost zone of the helium core (zone 5) is characterised by a C/O ratio~$>1$, and is conducive to the synthesis of carbon and silicon carbide dust. 

The mass of the various dust components as a function of post-explosion time is illustrated in Figure \ref{fig1}, where the mass is summed over all zones and corresponds to grains with size larger than 10~\AA. We consider that below this radius, the grains are molecular dust clusters. Silicates with forsterite stoichiometry (\fos) starts forming mainly in zone 1B as early as 300 days post-explosion. Around day 600, the mass of forsterite is boosted by the formation of grains in zone 2. The formation of alumina in zones 2 and 3 adds up to the total dust mass in the oxygen core after day 700. Carbon grains form at late time ($t > 1000$ days) because the formation of carbon clusters is hampered by the presence of high quantities of \hep\ in zone 5 (SC13). Overall, the condensation follows the trends derived by SC13 for the nucleation of small dust clusters in the ejecta. Dust formation occurs by following several dust synthesis events at different times and in different zones of the ejecta, and results in a gradual build-up of the dust mass over a time span of a few years after the explosion. The total mass of dust at day 400 is $\sim1 \times 10^{-5}$~\Ms\ and gradually grow to reach 0.035~\Ms\ four years after the outburst. The general trend in dust production indicates a high efficiency of condensation of the gas-phase precursors that amounts to $\sim$~99~\%. 

To explore how dust grains are distributed over size, we define for a specific dust type $i$ the grain size distribution $f_i(a)$ as

%----- eq 8 ---------
\begin{equation}
\label{eq8}
N_{tot, i}(a) = \sum_{a} f_i(a)\times \Delta{a}, 
\end{equation}
%----------------------

where $N_{tot,i}(a)$ is the total number of grains with radius $a$ summed over all zones. The quantity $N_{tot, i}(a)$ is calculated from the number density $n_i(a)$ of grains with radius $a$ and by assuming spherical symmetry for the various ejecta zones. The size distribution $f_i(a)$ has thus the units of $N_{tot,i}(a)$ {~\AA}$^{-1}$. The size distributions for the various dust types as a function of post-explosion time are presented in Figure \ref{fig2}.

%------------ fig3 ----------------------
\begin{figure}
   \includegraphics[width=\columnwidth]{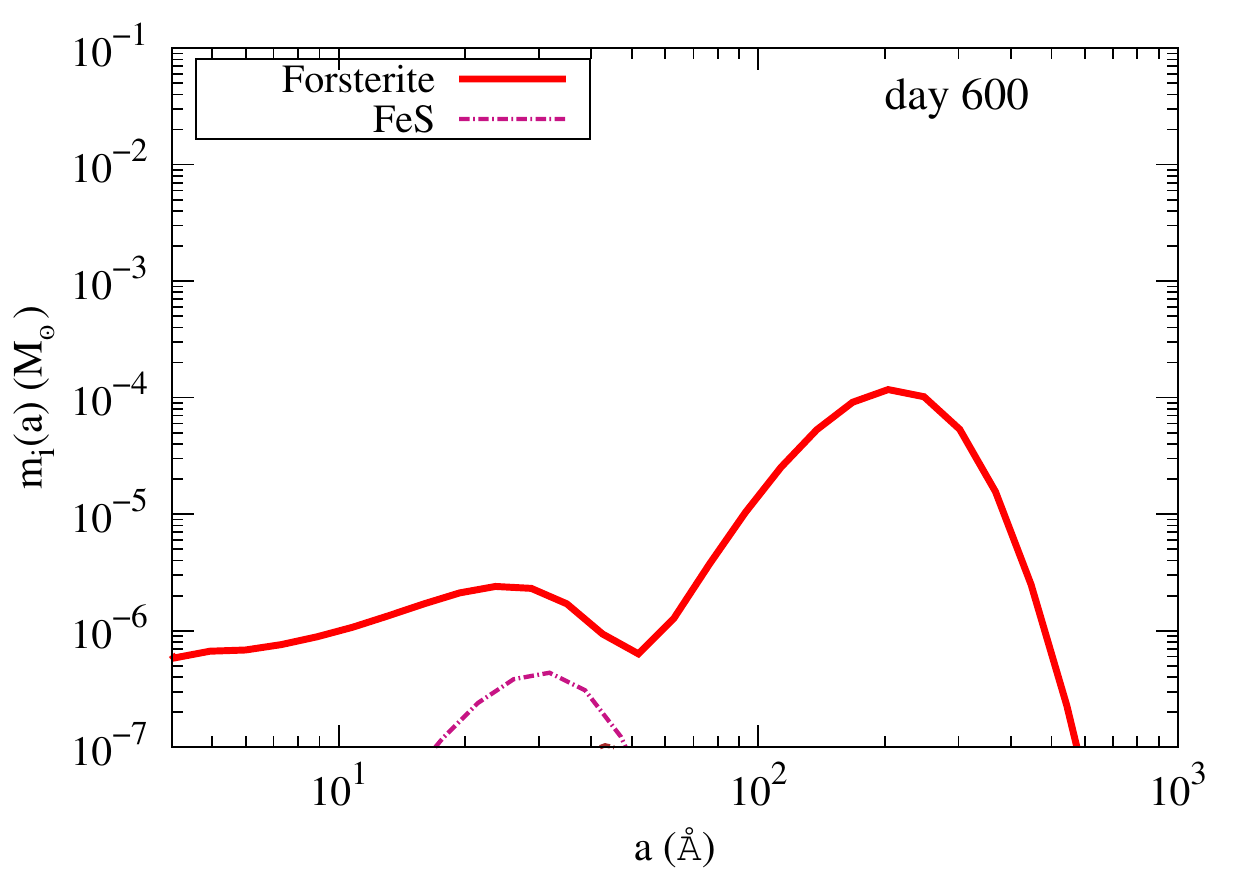}
    \includegraphics[width=\columnwidth]{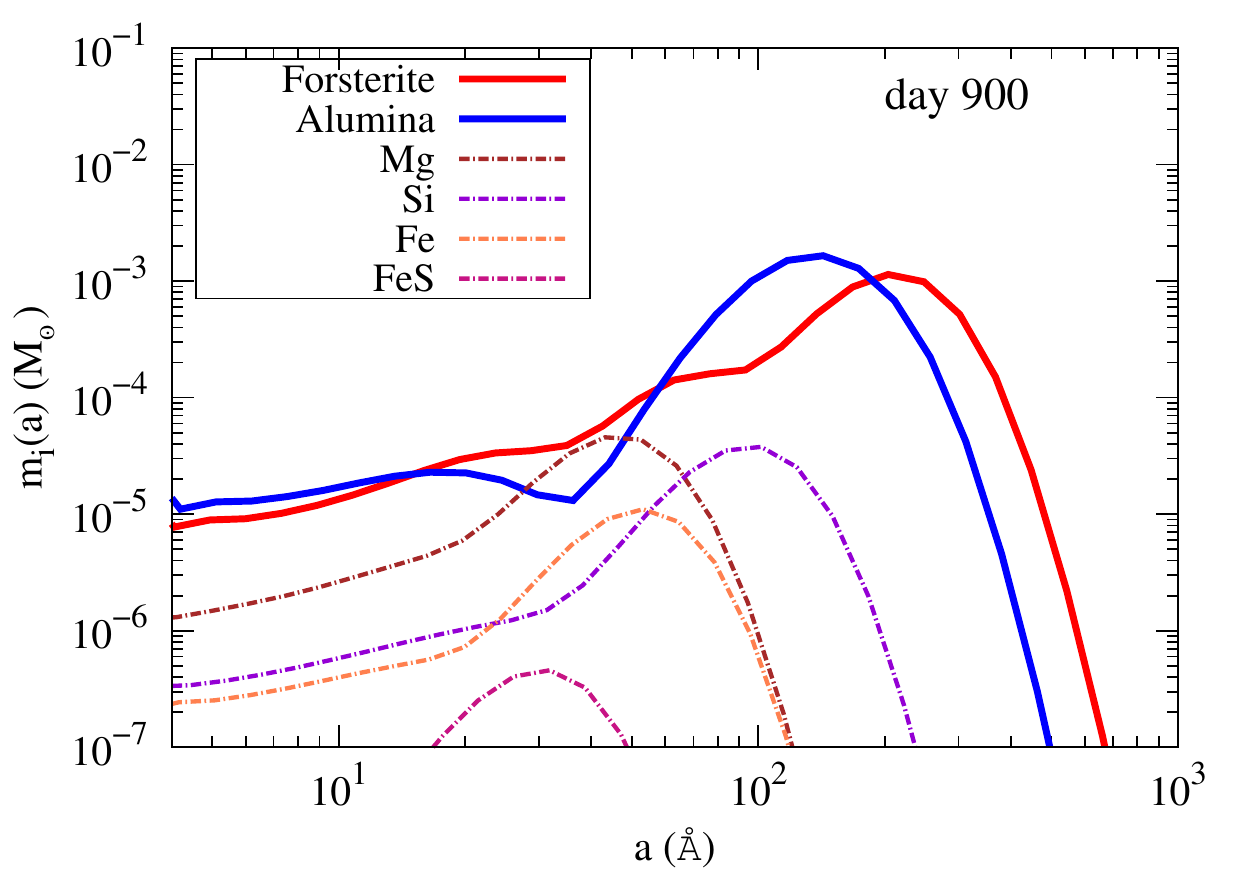} 
   \includegraphics[width=\columnwidth]{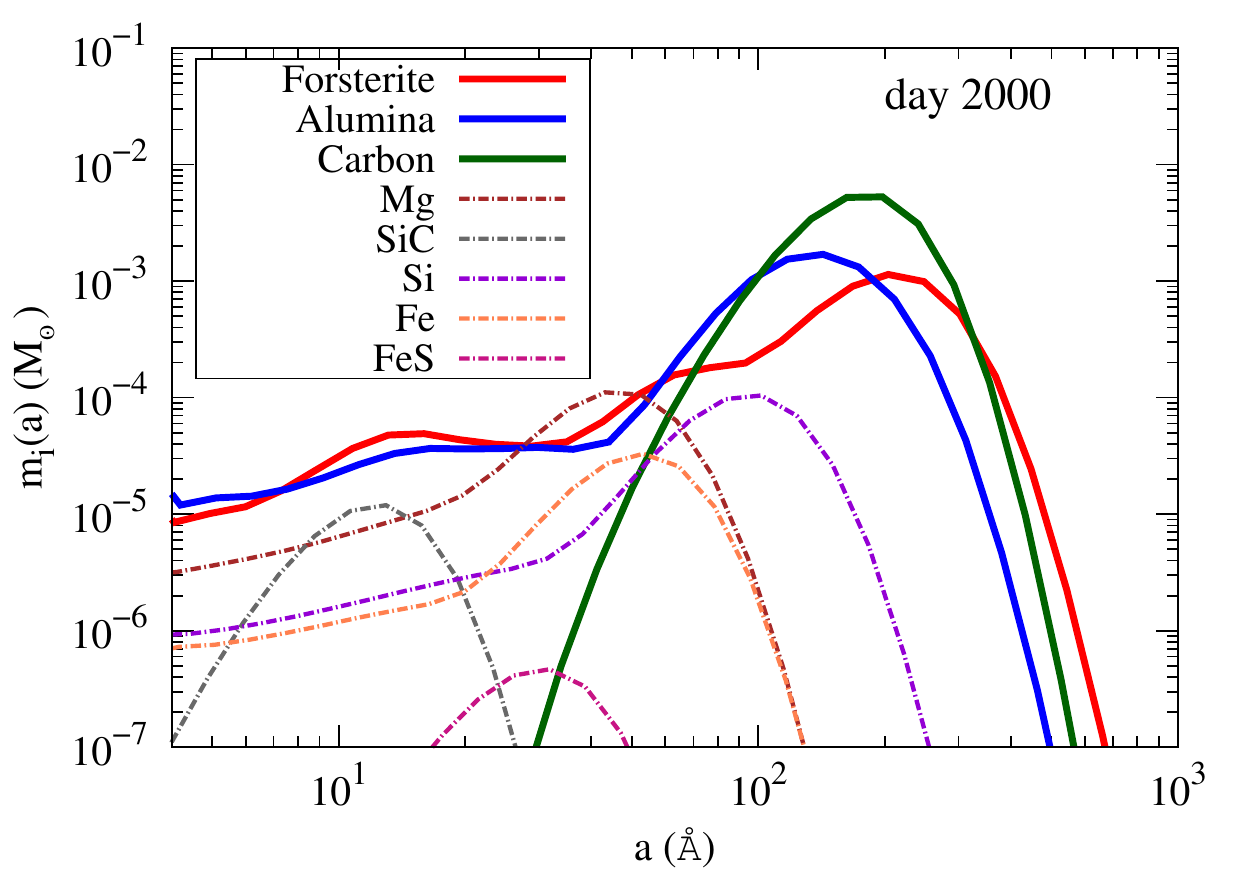} 
 \caption{Dust mass distributions (in~\Ms) as a function of grain radius for the various dust types formed in the ejecta at 600 days (top), 900 days (middle), and 2000 days (bottom) after explosion.}
 \label{fig3}
\end{figure}
%--------------------------------------
 
 %--------------- Table 5 --------------
\begin{table*}
\centering
\caption{Dust masses at different post-explosion epochs for the 15~\Ms\ progenitor, standard case, and the mass fraction of each dust component $x_d$ (in \%) with respect to total dust mass. The peak size $a_{peak}$ of each size distribution is also indicated.}
\label{tab5}
\begin{tabular}{l c c c c c c c c c c}
\hline \hline
Dust type & \multicolumn{7}{c}{Post-explosion time (days)} & $x_d$  & $a_{peak}$ (~\AA) \\
\hline
 & 500 & 700 & 900 & 1100 &1200 & 1500 & 2000 & & \\
\hline
Forsterite & 1.1(-5)& 4.2(-3)&  5.3(-3)& 5.5(-3) & 5.5(-3) &  5.6(-3) & 5.6(-3) & 15.8 & 64; 168\\
Alumina & - & 6.1(-6)  & 7.4(-3) & 7.6(-3) & 7.6(-3) &7.7(-3) & 7.7(-3) & 21.8 &96\\
Carbon & -& - & - & 2.0(-2) & 2.0(-2) & 2.0(-2)& 2.1(-2)& 59.7 & 134 \\
Pure Magnesium & -& 2.5(-6) & 1.9(-4) & 3.2(-4) & 3.6(-4)& 4.4(-4) &5.0(-4) &1.4 &29 \\
Pure Silicon & -& - & 1.5(-4) & 2.7(-4)& 3.1(-4)&3.8(-4) & 4.4(-4) & 1.2 & 0.69  \\
Pure Iron & -& - & 4.1(-5) & 9.1(-5) & 9.4(-5) &1.2(-4)&1.4(-4) & 0.4 &36 \\
Silicon Carbide & -& - & - & - & 6.1(-6) &2.6(-5)& 3.4(-5) & 0.1 &9 \\
Iron Sulphide & 1.7(-6) & 1.8(-6) & 1.8(-6) & 1.8(-6) & 1.8(-6) &1.8(-6)&1.8(-6) & 0.05& 21 \\
\hline
Total & 1.3(-5) & 4.3(-3) & 0.013 & 0.034 & 0.034& 0.035& 0.035 & 100 & \\
\hline

\end{tabular}
\end{table*}
%---------------------------------------------------

Small grains of silicate dust are produced at day 300 in the inner zone 1B. The average grain size remains small ($\sim 50$~\AA) at day 500 as well as the mass produced, because zone 1B has a small mass compared to zone 2. The forsterite dimers efficiently starts forming larger grains at $\sim$ day 700 from zone 2 where the gas temperature is $\sim$ 1000 K, and the peak of the distribution curve shifts from 50~\AA\ to 100~\AA. The decrease in the number of small grains around 20~\AA\ is attributed to the formation of larger grains. After 1000 days, the gas density is quite low and no new nucleation seeds (small clusters) form. The concentration of large grains ($a >200$~\AA) is also low, and their growth is stopped. However small grains still participate in the coagulation process by replenishing the grain population with size 10~\AA\ $<a<$ 100~\AA, albeit with a lower efficiency. When coagulation is completed at day 2000, the silicate grain size distribution has a peak of 170~\AA\ for the largest grain population, and most of the large grains come from zone 2.

Alumina is the second dust component to form in the O-rich core at $\sim$ day 700 after explosion. The alumina grains undergo very fast and efficient condensation to form large grains peaking at $\sim$ 60~\AA\ at day 900, when the gas temperature is in the range $600-800$ K. As shown in Figure 13 of SC13, most atomic aluminium present in the O-rich core gets locked up in molecules and dust clusters, hence no new \alu\ dimers are formed after day 900 and a growth trend similar to that of forsterite applies to alumina. The alumina dust growth results in the production of large grains peaking around 100~\AA\ at day 2000. 

The formation of carbon dust in the outermost ejecta zone is strongly affected by the presence of \hep. The synthesis of stable C$_{28}$ cages occurs as late as 1050 days after outburst once the abundance of \hep\ ions has decreased to negligible values  (SC13). Then the carbon chains, rings and cages form simultaneously with high abundances along with the efficient condensation of the cage C$_{28}$ in carbon grains. This results in a sudden high concentration of large carbon grains with peak radius of $\sim$ 100~\AA\ at day 1200. The rate of coagulation is directly proportional to the thermal velocities and hence the gas temperatures, as seen from Equation \ref{eq3}. Because of the low gas temperatures ($\sim$ 300 K) and densities at $t > 1200$ days, the condensation process becomes less efficient. However, owing to a high abundance of carbon cages in the gas phase, the condensation of grains proceeds even at these low gas temperatures. The carbon grain size distribution does not vary after day 1500, and shows a peak for large grains at $a \sim$ 150~\AA. 

Apart from the previous three prevalent dust components, iron sulphide, pure metal, and silicon carbide grains also condense in the ejecta. Iron sulphide, FeS, forms at day 300 in the innermost zone, zone 1A, and reaches its final mass and size distribution at day 500, with a peak of the distribution curve that lies at 25~\AA. Pure silicon and iron grains start forming at day 700  in zone 1A, where most of the silicon is locked in the molecule SiS and in atomic form (SC13). Iron is mainly in atomic form in this zone but some pure iron clusters form after day 700. The size distribution of Si grains is the most extended of all pure metal dust grains, and peaks at $\sim$ 70~\AA, although the overall grain population remains small because of the modest amounts of pure Si clusters formed in the gas phase. A similar scenario applies to pure iron grains, for which the distribution peaks at 40~\AA\ at day 2000. The pure magnesium grains are synthesised from the Mg$_4$ clusters formed in the O-rich zones at day 600. The size distribution reaches its final shape at day 1200, with a peak at $\sim 30$ \AA, and a small population of grains with sizes over 120~\AA. Silicon carbide forms in the He/C-rich zone 5 after day 1100, and captures the available silicon and the carbon left over from the condensation of carbon dust. The low abundance of SiC clusters combined to the low gas temperature and densities at the epoch of its formation result in small masses of SiC grains characterised by small sizes in the range $5-11$~\AA. The various dust masses versus post-explosion time and the radius $a_{peak}$ where the size distributions peak are summarised in Table \ref{tab5}. 

In Figure \ref{fig2} (bottom-right panel), the total dust size distribution is shown at day 900 and 2000, along with a dust size distribution for a similar dust mass following a power-law of exponent -3.5, and the Mathis-Rumpl-Nordsieck power law distribution (exponent -3.5) for grains with radius $a$ in the range $50-2500$ \AA, characteristic of interstellar dust (\cite{mat77}). The total size distribution is the total number of grains at a particular size $a$, calculated by summing the $N_{tot,i} (a)$ for all dust types $i$ over a small size interval centred on $a$. The size distributions of the dust produced by SNe do not follow a power law, for both single dust types and the total dust distribution. Our size distributions have less small grains with radius $a < 30$~\AA, and are skewed towards large grains in the size range $40-1000$~\AA, compared with the a power law distribution of any exponent. Furthermore, our total number of grains are roughly one order of magnitude larger compared to the MRN profile around 200~\AA. This result clearly indicates that the estimated dust mass that survives sputtering by shocks in the remnant phase may be incorrect if a MRN size distribution is assumed for the unshocked dust grains present in the SNR phase.

The dust mass distributions as a function of grain size are shown in Figure \ref{fig3} at 600, 900, and 2000 days after explosion, and are derived as follows: 
%The total mass of grains of a specific chemical type and with radius $a$ is derived 
%from the grain number density $n(a)$, the grain mass according to stoichiometry, and the molecular weight and the mass of each ejecta zone. The quantity $N_{tot}(a)$ is then estimated using the densities of individual components. 
for a specific chemical dust type $i$ and epoch, the mass fraction $x_{m,i}(a)$ of grains with radius $a$ is given by
%----- eq 9 ---------
\begin{equation}
\label{eq9}
x_{m,i}(a) = \frac{n_i(a) \times v_n(a)}{\sum_{a} n_i(a) \times v_n(a)}
\end{equation}
%----------------------
where $v_n(a)$ is the volume of the $n^{th}$ bin that corresponds to radius $a$, and $n_i(a)$ is the number density of grains with radius $a$. The total mass of dust $m_i(a)$ with radius $a$ is then given by
%----- eq 10 ---------
\begin{equation}
\label{eq10}
m_i(a) = x_{m,i}(a)\times M_{tot,i} = N_{tot,i}(a) \times {v_n(a) \times \rho_{d,i}}
\end{equation}
%----------------------
where $M_{tot,i}$ is the total mass of condensed dust of type $i$ in the ejecta, and $\rho_{d,i}$ the density of the dust material.

We see that the dust mass quickly grows with time and mainly resides in a population of large grains with sizes that range from 60~\AA\ to 700~\AA. Once coagulation has stopped, most of the dust mass is distributed in populations of carbon, forsterite, and alumina grains, while pure metals, FeS, and SiC grains do not significantly contribute to the total dust mass. For the standard case with 15~\Ms\ progenitor, carbon dust dominates with a final mass amounting to $\sim 0.02$~\Ms, while the alumina and forsterite masses are $0.008$~\Ms, and $0.006$~\Ms, respectively, as summarised in Table~\ref{tab5}. 

More generally, about 99\% of the dust clusters formed in the gas phase efficiently condense to form dust, and only a small fraction remains in the gas phase. Coagulation is thus very efficient as soon as dust molecular clusters form in the ejecta gas. However, we see from Figure \ref{fig1} that the dust size distributions are strongly time-dependent. The overall dust formation process in the ejecta results in a gradual build-up of various dust grain populations in the different ejecta zones. Therefore, the dust observed at IR wavelengths at post-explosion day 500 differs in type, mass, and size distribution, from the dust observed at submm wavelengths. This late dust is the outcome of a complete series of nucleation and coagulation events over a time span of $\sim$ 5 years after outburst. 

\subsection{15~\Ms\ progenitor with \Ni\ = 0.01~\Ms} 
\label{lowni}

% ----------------------- Figure 4 --------------------------------
\begin{figure}
\includegraphics[width=\columnwidth]{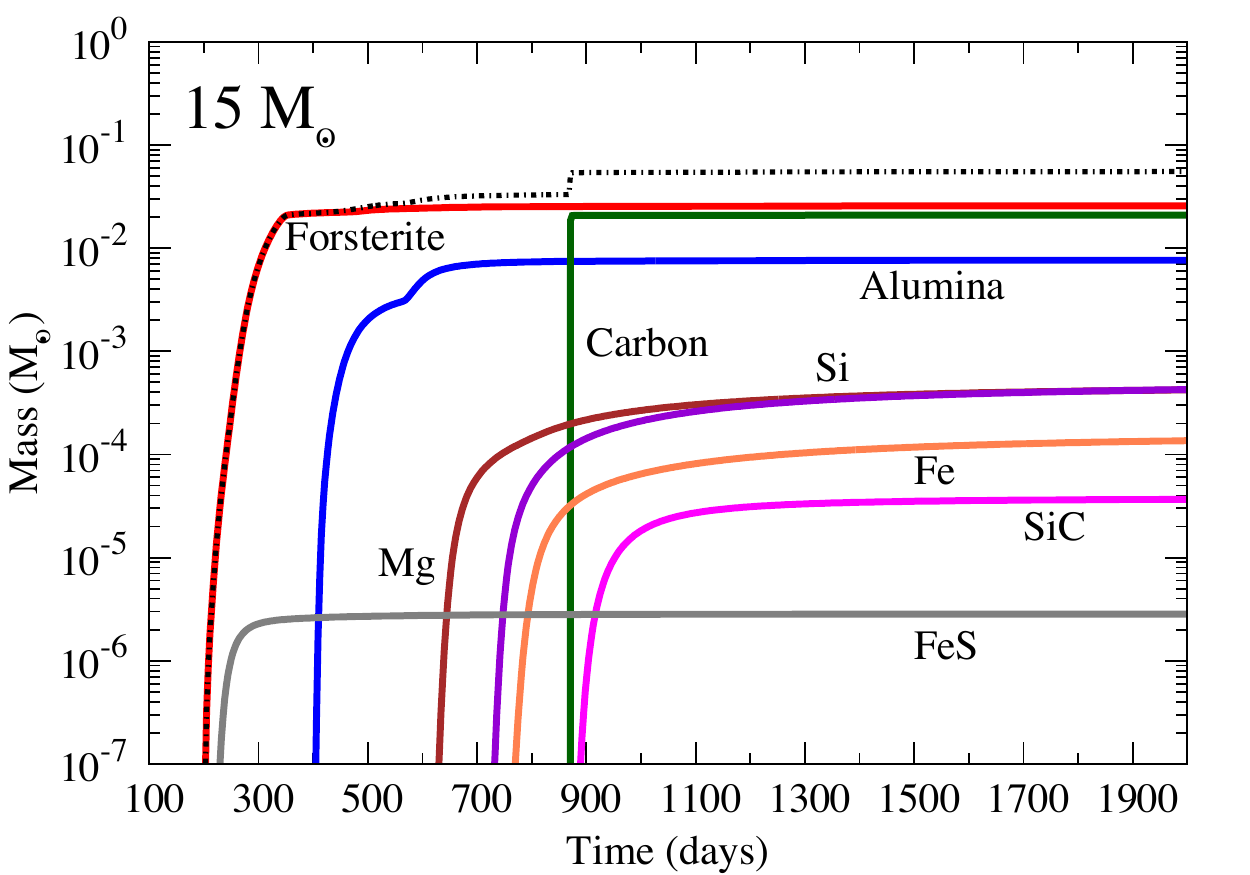}
\includegraphics[width=\columnwidth]{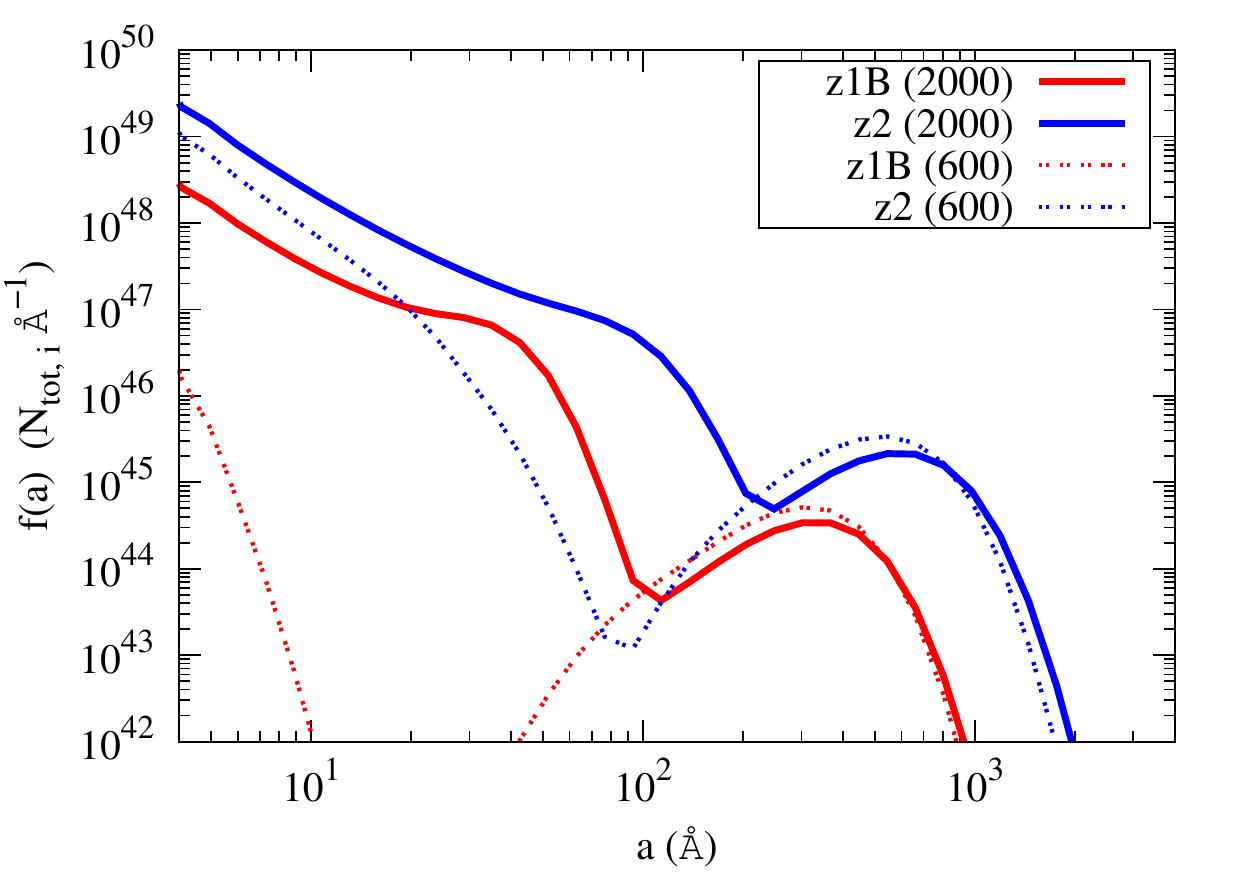}
\includegraphics[width=\columnwidth]{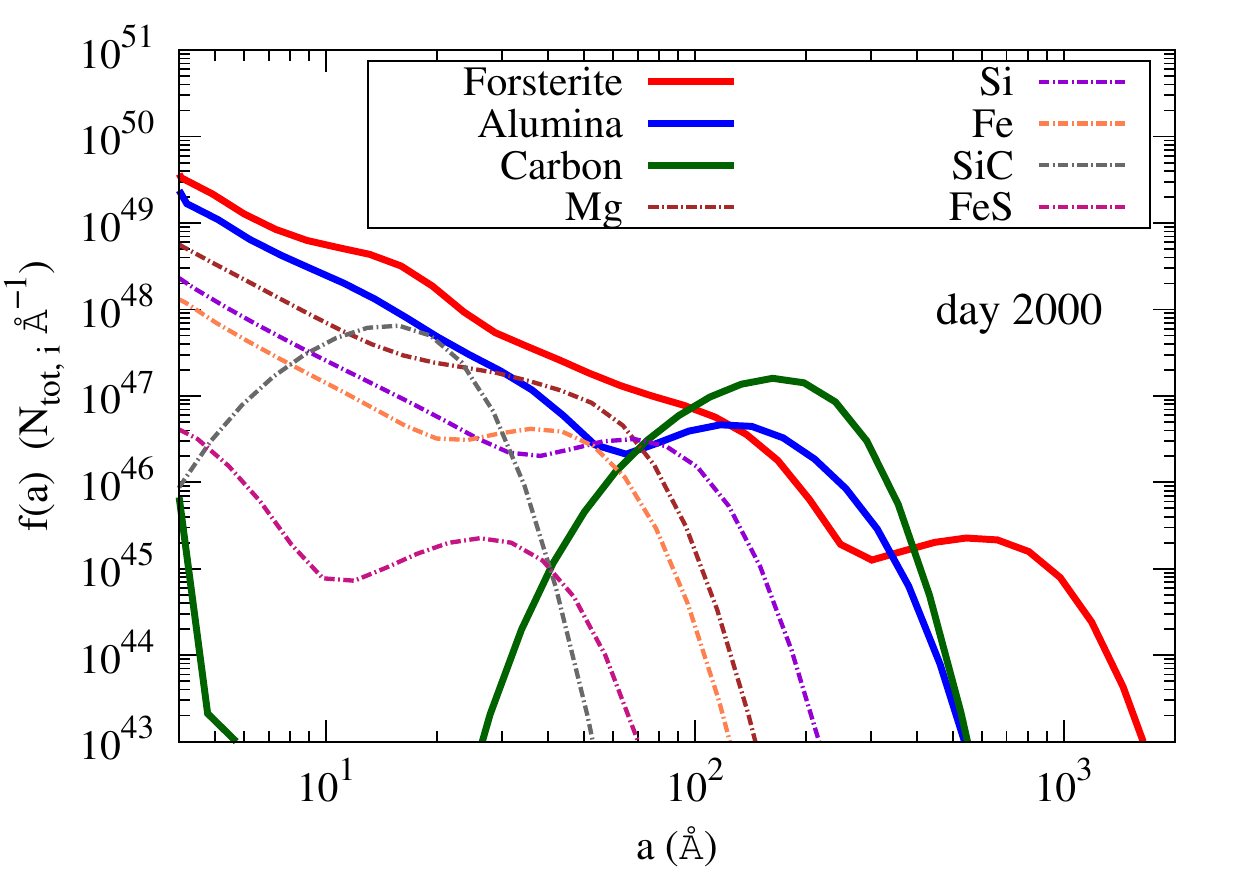}
\caption{Results for the homogeneous SN ejecta with 15~\Ms\ progenitor and \Ni $= 0.01$~\Ms. Top:~The total dust mass versus post-explosion time; Middle: The dust size distributions for forsterite grains in zone 1B and 2 at day 600 and 2000; Bottom: The dust size distributions at day 2000.}
\label{fig4}
\end{figure}
%-------------------------------------------------------------------

Several SN explosions have stellar progenitors in the mass range $8-15$~\Ms\ and produce \Ni\ mass between 0.01 and 0.02~\Ms. The affect of \Ni\ on the gas phase chemistry has been assessed and discussed in SC13. We extend our analysis to study the effect of a small \Ni\ mass on the dust condensation process and the resulting dust masses and size distributions. The less efficient deposition of \grays\ energy due to a smaller \Ni\ mass will decrease the ejecta gas temperature. However, we do not consider any change in the ejecta physical conditions caused by decreasing the \Ni\ mass. The physical conditions and initial chemical compositions for each ejecta zones are then given in Table \ref{tab1} and correspond to the standard case, but the \Ni\ mass has been changed to $0.01$~\Ms. The effect on the nucleation of dust clusters is that described in SC13. However, a low \Ni\ also affects the dust mass and size distribution, as shown in Figure \ref{fig4}, where the mass of dust summed over all ejecta zones is displayed, along with the forsterite grain size distributions versus formation zones and post-explosion time, and the size distributions for the various dust components. 

Forsterite, alumina and carbon dust are produced in zones 1B, 2 and 5, respectively, and these specific zones are characterised by high abundances of noble gases, which include argon, neon, and helium. The \gray~flux induced by the decay of \Ni\ is lower owing to the lower \Ni\ mass. Therefore, the abundances of \arp, \nep, and \hep\ ions produced by the degrading of \grays~are lower and dust clusters form early, starting at $\sim$ day 250 (SC13). The coagulation of dust clusters occurs at high gas density from a larger mass of clusters than in the standard case. This efficient coagulation ensues the production of large grains. Inspection of Figure \ref{fig4} (middle panel) shows that for forsterite, the formation of large grains is already completed in zones 1B and 2 at day 600, with almost no variation of these large grain populations at later time. The peaks in the size distribution are from grains produced in zone 2, and correspond to 130~\AA\ and 550~\AA. For alumina, the distribution peaks around 120~\AA\, and for carbon, the peak is located at 160~\AA. Pure silicon and iron dust have similar distributions than for the standard case because both dust types form in zone 1A, which is deprived of inert gas elements. 

The final dust mass produced in this ejecta is 0.055~\Ms, which represents an increment of 57 \% compared to the value for the standard case. Therefore, for an equal progenitor mass, the ejecta characterised by a low mass of produced \Ni\ forms a larger quantity of dust grains as early as day 300 and with larger radii than in the case of a high \Ni\ mass SN counterpart. 

%-------- table 6 ---------
\begin{table}
\centering
\caption{Dust types and masses produced in the ejecta of a 15~\Ms\ progenitor with M(\Ni) $= 0.01$~\Ms, and the mass fraction of each dust component $x_d$ (in \%) with respect to total dust mass. The peak positions in the grain size distribution are also listed.}
\label{tab6}
\begin{tabular}{l c c c}
\hline
\hline

Dust type & Mass (~\Ms) & $x_d$  &$a_{peak}$ (\AA) \\
 
\hline

Forsterite &  2.5(-2) & 45.5&131; 543 \\
Carbon &  2.1(-2) &38.2 &162 \\
Alumina  & 7.7(-3) & 14.0&117 \\
Pure Silicon & 4.4(-4) &0.80 &69 \\
Pure Magnesium  &  4.3(-4) &0.78 &53 \\
Pure Iron &  1.4(-4) & 0.25&36 \\
Silicon Carbide  &  8.8(-5) & 0.16&16 \\
Iron Sulphide  &  2.8(-6) & 0.005& 26\\
\hline0.00

Total & 0.055 & 100&\\
\hline 
\end{tabular}
\end{table}
%---------------------------------

%---------------- Figure 5 -------------------------
\begin{figure} 
\centering
\includegraphics[width=\columnwidth]{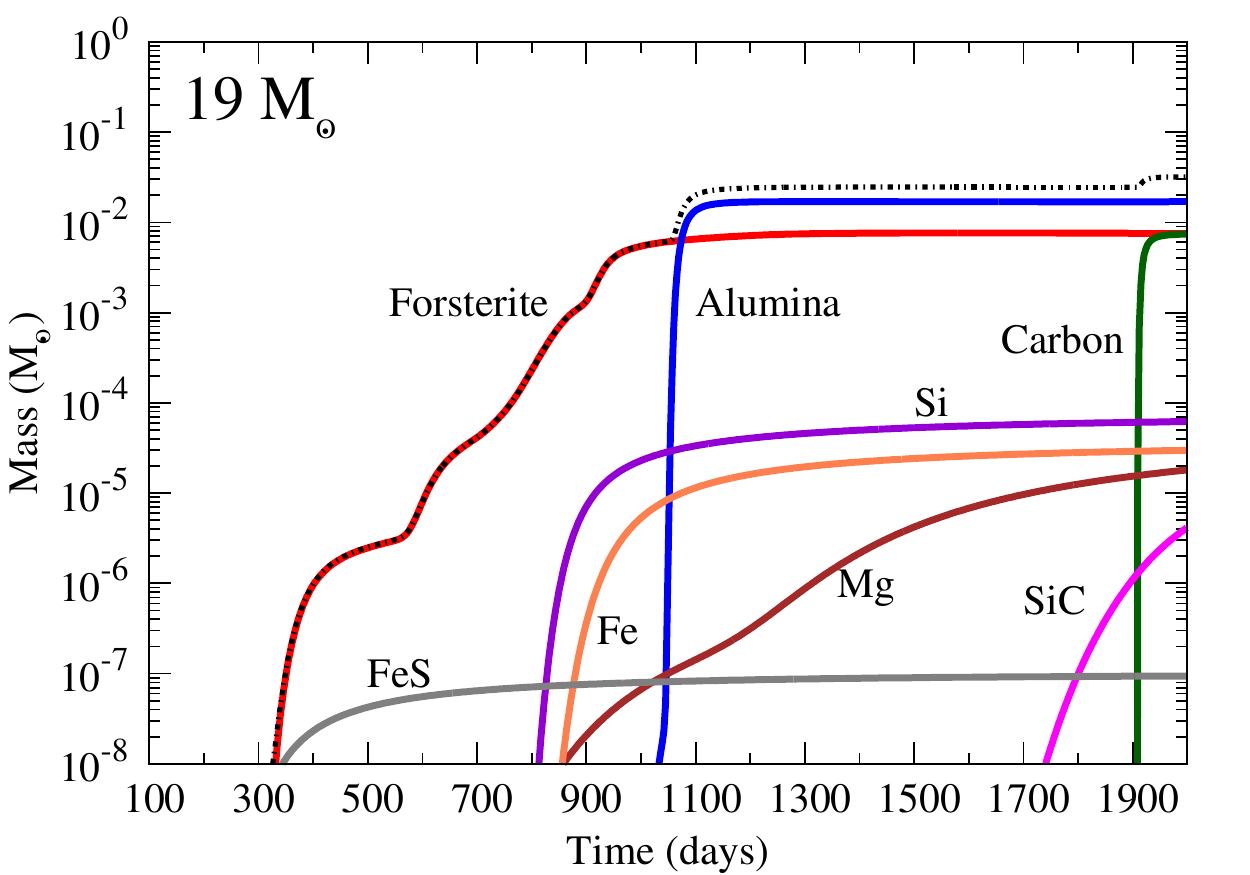}
\includegraphics[width=\columnwidth]{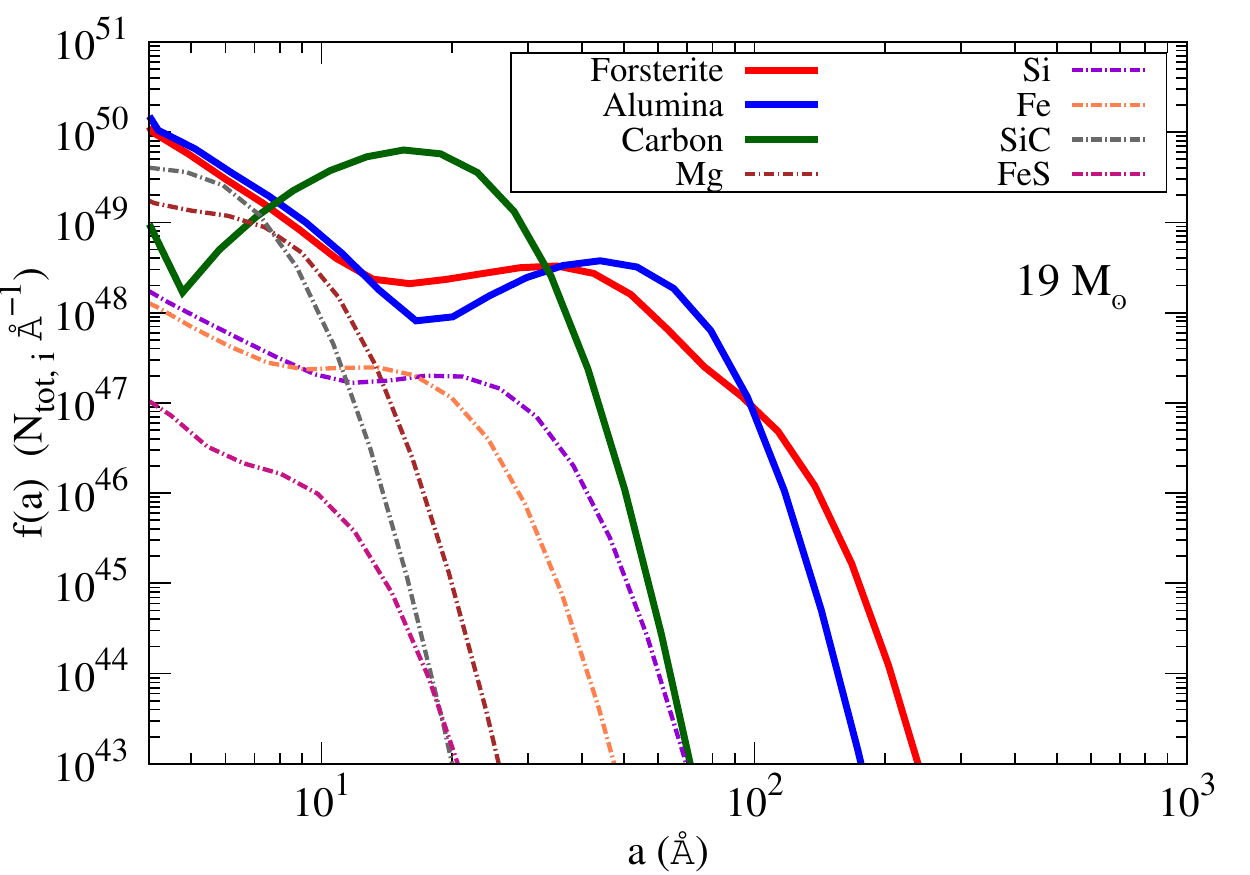} 
\caption{Top: dust mass as a function of post-explosion time and dust types for the 19~\Ms\ progenitor with homogeneous ejecta; Bottom: dust size distributions as function of dust type at day 2000. }
\label{fig5}
\end{figure}
%-----------------------
\subsection{19~\Ms\ model: homogenous and clumpy ejecta} 
\label{19ms}

For the SN originating from the explosion of a 19~\Ms\ stellar progenitor, we consider two cases: 1) a homogeneous, stratified ejecta, and 2) a clumpy, stratified ejecta whose model is given in Table \ref{tab3}. For the homogeneous case, we consider an initial gas density of $4.4 \times 10^{-12}$ g \cmc\ for all mass zones at day 100. This value is a factor 2.5 smaller than that considered in SC13, and was arbitrary chosen to offer a better match between the modelled SiO masses for the clumpy case and those derived from observations - see \S\ \ref{clump}. As explained in \S\ \ref{sec2}, the resulting gas number density per zone in the clumpy case is corrected by the volume filling factor derived by Jerkstrand et al. (2011).  

\subsubsection{Homogeneous} 

Results on the dust mass and the grain sizes distributions for the various dust types synthesised in the ejecta are presented in Figure \ref{fig5}. The final mass of dust formed stems from various dust production events in ejecta zones at different time, similar to what happened in our 15~\Ms\ progenitor case. However, the slighlty lower number densities for each ejecta zone at day 100 lead to a delayed formation of forsterite clusters in zones 1B and 2, as seen in Figure \ref{fig5}, compared to the results presented in Figure 10 of SC13 for a similar progenitor mass. 
Iron sulphide, pure iron and silicon dust first condense in zone 1A, followed by forsterite in zone 1B, forsterite, alumina and pure magnesium in zones $2 -4$, and finally, carbon and silicon carbide dust in zone 5. The condensation of dust clusters is efficient as 96 \% of the clusters initially formed in the gas phase condense into dust grains. The final dust mass at day 2000 amounts to 0.032~\Ms, and details on the various dust populations, their masses and peak radii are given in Table~\ref{tab7}.

The homogeneous ejecta forms medium-size grains, with distribution peak sizes for alumina, and forsterite ranging between $30-70$~\AA, while the synthesis of carbon dust results in a population of small grains, that peaks around 20~\AA. The largest grains correspond to the forsterite grains formed in zone 1B before day 600, as these grains had more time to grow than the forsterite grain population produced in zone 2. The distribution resembles that for the 15~\Ms\ progenitor, except for the smaller contribution from carbon dust to the total dust budget, as seen from Table \ref{tab7}. 

\subsubsection{Clumpy}
\label{clump}
%--------------------- Figure 6 ----------------------------
\begin{figure*}
 \centering
\includegraphics[width=\columnwidth]{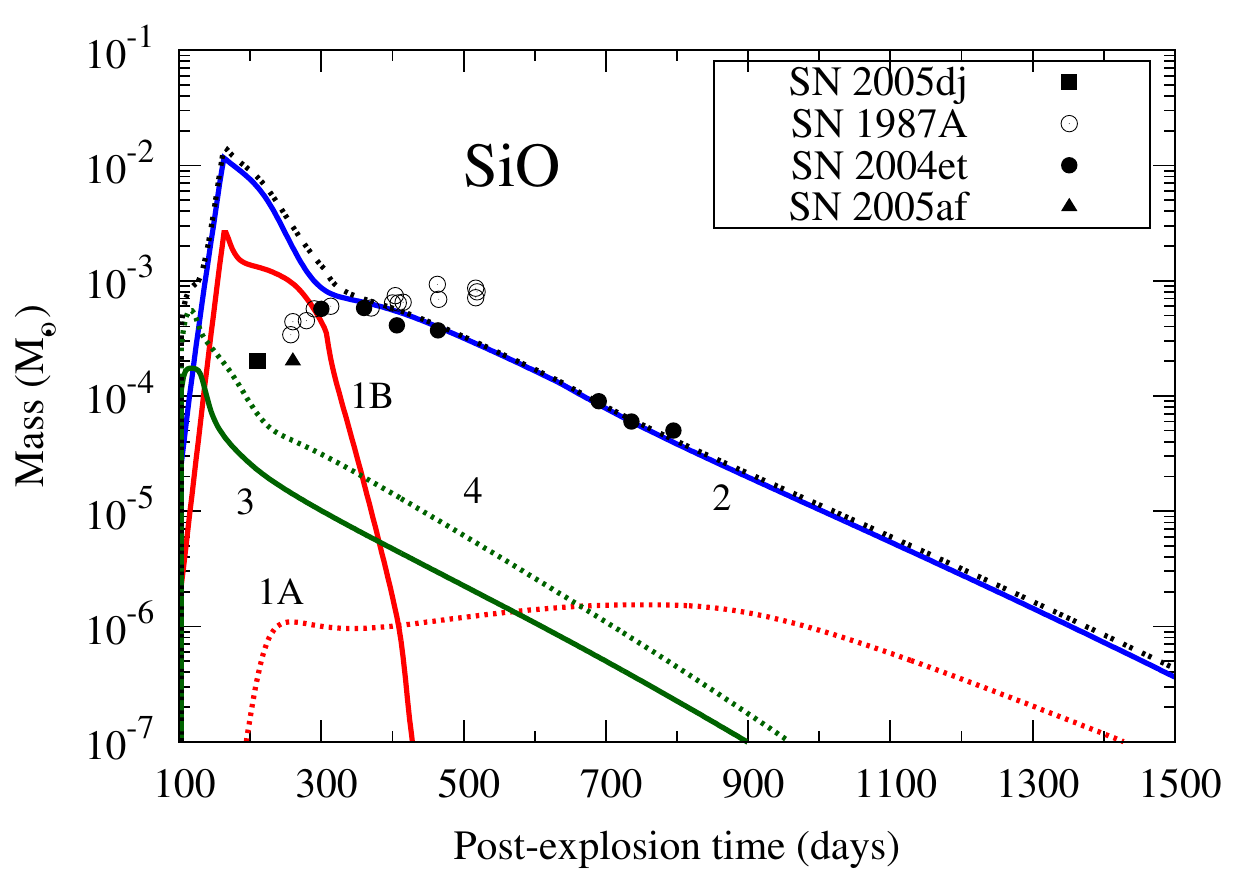}
\includegraphics[width=\columnwidth]{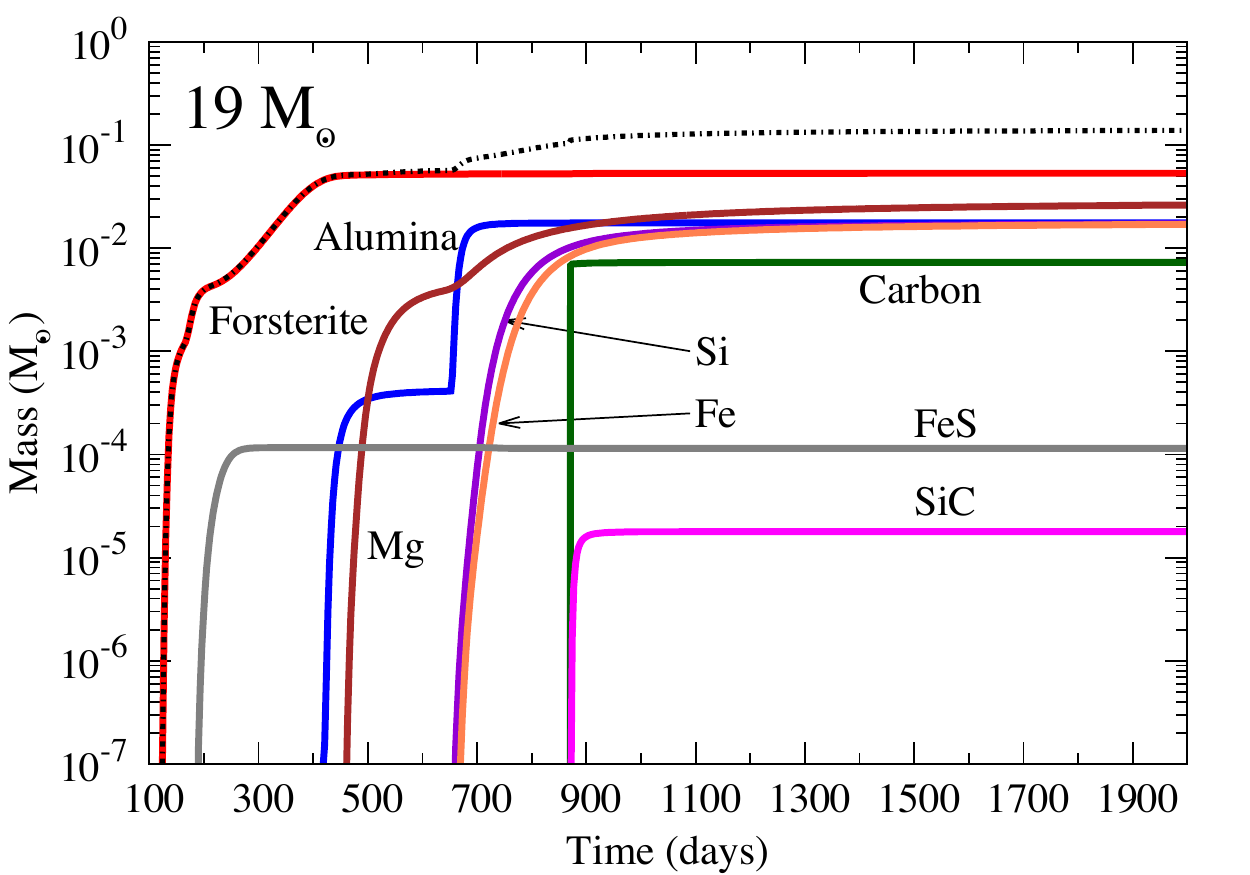} 
\includegraphics[width=\columnwidth]{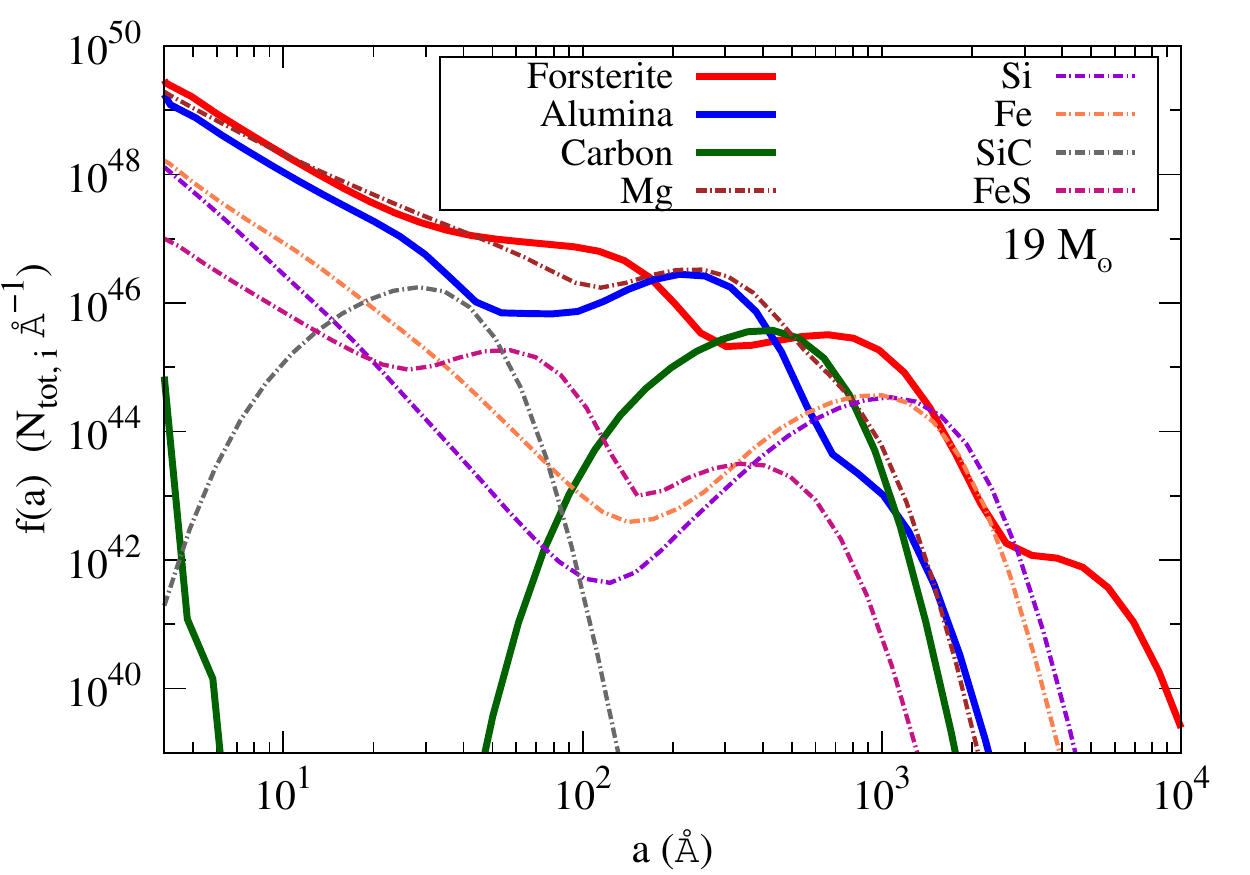} 
\includegraphics[width=\columnwidth]{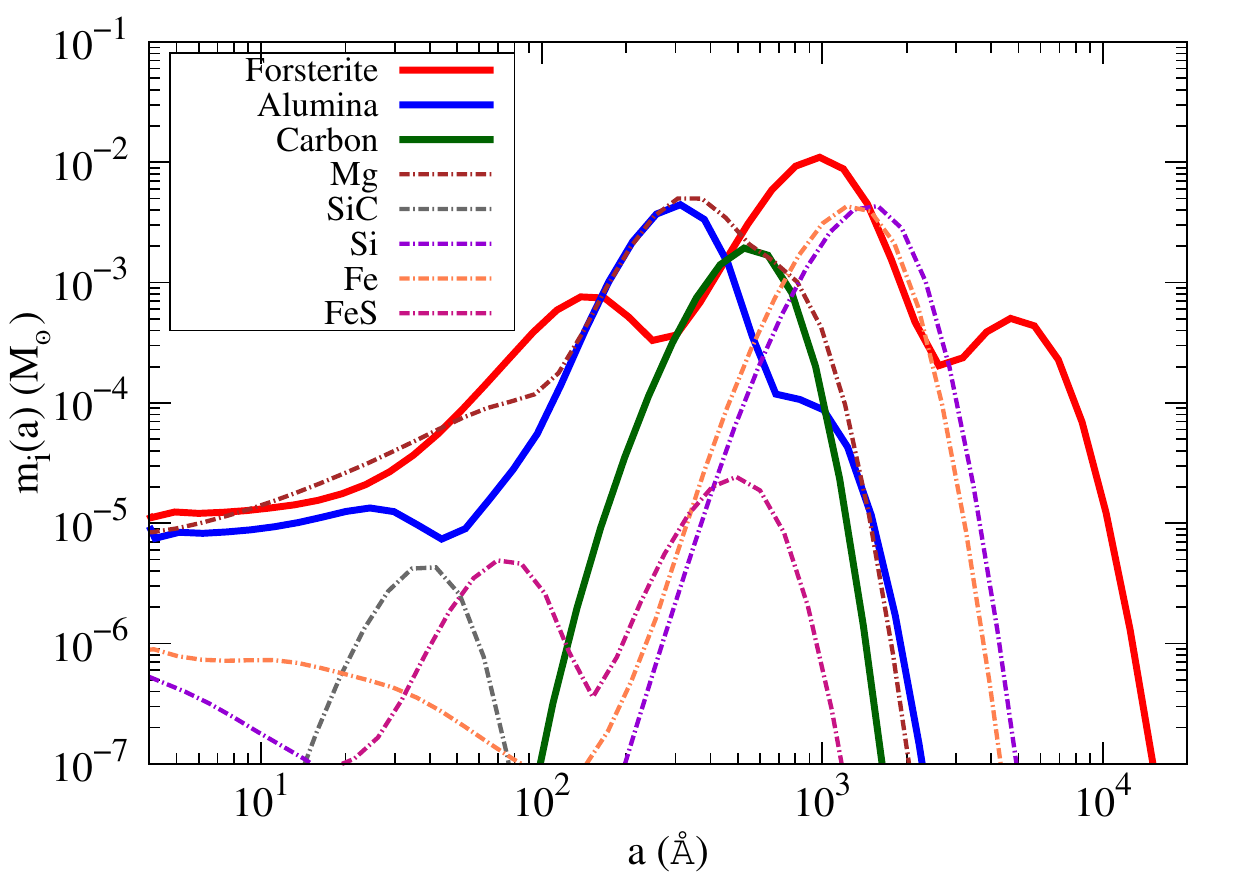} 
 \caption{Results for the 19~\Ms\ clumpy model: Top-left: SiO mass as a function of post-explosion time. Observational data for several supernovae are given as symbols. Top-right: dust mass as a function of post-explosion time and dust type. Bottom-left: grain size distributions of the various dust components at day 2000. Bottom-right: Mass distributions of the various dust components at day 2000.}
 \label{fig6}
\end{figure*}

%------------------------------------------------------------
%--------------- Table 7 --------------
\begin{table*}
\caption{Dust masses formed at day 2000 post-explosion as a function of dust type and ejecta zone for the homogeneous and clumpy ejecta of a 19~\Ms\ stellar progenitor. Also shown is the mass fraction of each dust component $x_d$ (in \%) with respect to total dust mass, and the $a_{peak}$ radius for the various dust distributions.}
\label{tab7}
\centering
\begin{tabular}{l c c c c c c c r c}
\hline \hline
%Dust type & \multicolumn{7}{c}{Time post-explosion (days)} \\
%\hline
\multicolumn{10}{c}{HOMOGENEOUS} \\
\hline
 Dust type & Zone 1A & Zone 1B & Zone 2 & Zone 3 & Zone 4 & Zone 5 & Total & $x_d$ & $a_{peak}$ (~\AA) \\
\hline
Forsterite & - & 2.3(-3)& 5.1(-3)&  9.1(-5)& 1.0(-4) &  - & 7.6(-3) & 23.8  & 35, 63\\
Alumina & - & - & 1.7(-2) & 3.9(-6) & 6.5(-5) & - & 1.7(-2) & 53.0 & 44\\
Carbon & -& - & - & - & - &7.5(-3) & 7.5(-3) & 23.4 & 16 \\
Pure Magnesium & -& - & 1.7(-5) & 2.1(-7) & 5.4(-7)& - & 1.8(-5) & 0.06  & 8 \\
Pure Silicon & 6.2(-5)& - & - & -&- & - & 6.2(-5) & 0.19  & 18 \\
Pure Iron & 3.0(-5)& - & - & -&- & - & 3.0(-5) & 0.09 & 14 \\
Silicon Carbide & -& - & - & - & - & 3.7(-6) & 4.1(-6) & 0.01 & 7 \\
Iron Sulphide & 9.4(-8) & - & - & -&- & - & 9.4(-8)  & 3(-4)  & 8 \\
\hline
Total & 9.2(-5) & 2.3(-3) & 2.2(-2) & 9.5(-5) & 1.7(-4) & 7.5(-3)& 0.032 & 100 &\\
\hline
\multicolumn{10}{c}{CLUMPY} \\
\hline
 Dust type & Zone 1A & Zone 1B & Zone 2 & Zone 3 & Zone 4 & Zone 5 & Total & $x_d$  &  $a_{peak}$ (~\AA) \\
\hline
Forsterite & - & 2.4(-3) & 4.8(-2) &  5.4(-4) & 1.6(-3) & 3.0(-4) & 5.3(-2) & 38.4 & 77, 661, 3170 \\
Alumina & - & 4.0(-4)  & 1.7(-2) & 1.3(-5) & 6.3(-5) & - & 1.8(-2) & 13.0 & 211, 562\\
Carbon & -& - & - & - & - &7.3(-3) & 7.3(-3) & 5.3 & 435\\
Pure Magnesium & -& - & 2.1(-2) & 1.2(-3) & 4.0(-3)& - & 2.6(-2) & 18.8 & 252 \\
Pure Silicon & 1.7(-2)& - & - & -&- & - & 1.7(-2) & 12.3  & 1066\\
Pure Iron & 1.7(-2)& - & - & -&- & - & 1.7(-2) &12.3&  1003 \\
Silicon Carbide & -& - & - & - & - & 1.7(-5) & 1.7(-5) & 0.01& 29\\
Iron Sulphide & 1.1(-4) & - & - & -&- & - & 1.1(-4)  & 0.08 & 57, 334\\
\hline
Total & 3.4(-2) & 2.8(-3) & 8.6(-2) & 1.8(-3) & 5.6(-3) & 7.6(-3)& 0.138 & 100 &\\
\hline

\end{tabular}
\end{table*}
%--------------------------------------------------

Results for the clumpy ejecta are presented in Figure \ref{fig6}, where the SiO mass and the dust mass variation versus post-explosion time, the grains size distributions, and the grain mass distributions are shown. 

SC13 showed that SiO was a direct tracer of silicate dust condensation in the ejecta of supernovae. Here the decrease in SiO mass with time indeed agrees well with observational data for SNe with high progenitor masses, e.g., SN2004et (although the mass for this SN is still debated and may be lower than 20 \Ms, e.g., Jerkstrand et al. 2012), and SN1987A. Compared to the homogeneous ejecta, forsterite formation in Zone 1B occurs at early time and higher gas density. Furthermore, Zone 1B has the lowest filling factor among all the zones, which causes zone 1B to be the densest zone in the oxygen-rich core. This leads to the formation of a population of large grains, which peaks at $\sim 0.4$ \mic\ at day 2000, as seen in the dust size distributions (bottom-left panel). Forsterite also efficiently forms in zone 2 around day 400, and increases the total forsterite mass. This formation event is reflected in the size distribution by a population of large grains peaking at 660~\AA.

A similar scenario applies to alumina which forms essentially in zone 1B at $\sim$ day 450 and zone 2 at $\sim$ day 700. The alumina size distribution thus shows a peak around $\sim$ 240~\AA, which corresponds to the grains formed in zone 2, and a tail of large grains produced in zone 1B and with a size over 0.1 \mic. The formation of three populations of grains of pure silicon, pure iron, and iron sulphide pertains to zone 1A. Despite their late formation at day 650, pure silicon and Fe grain condensation is boosted compared to the homogeneous case because of the higher gas density in zone 1A. This leads to grain populations that peak around 0.12 \mic\ for both pure silicon and iron dust. This peak size is almost two orders of magnitude greater than in the homogeneous case. Grains of FeS condense at day 200, and thus grow over time to relatively large sizes ($\sim$ 400~\AA). However the final FeS dust mass remains low owing to a modest mass of FeS clusters that form in zone 1A. 

Finally, the outermost zone 5 forms carbon and SiC grains. The carbon grains form at day 870, a much earlier epoch than for the homogeneous case, where carbon forms at day 1350. This results in a size distribution dominated by larger grain sizes peaking at 520~\AA, while the carbon size distribution peaks at 70~\AA\ for the homogeneous ejecta.  

Dust condensation is strongly sensitive to the ejecta number density. A clumpy ejecta leads to the synthesis of all types of dust at early epochs, and increases the final dust mass, compared to the homogeneous case. The dust components can be further categorised in two groups. Clusters of alumina and amorphous carbon undergo a fast and efficient formation, but their final mass is controlled by the availability of atomic Al and C, respectively, in the production region. Therefore, the final mass of these dust populations is not significantly sensitive to any change in ejecta parameters, as seen from Table \ref{tab7}, for the typical ejecta gas parameters derived from SN explosion models. Conversely, the silicate and pure metal dust mass is strongly dependent of the gas conditions. Silicate clusters are formed through a complex silicon-based chemistry, whose efficiency strongly depends on the gas-phase parameters. The mass of silicate grains thus varies by a factor of 10 to 20 from the homogeneous to the clumpy case. As for pure metallic grains, the clusters essentially form in the gas phase through trimolecular association reactions that include the bath gas as a collider. These processes become very effective as the gas number density is raised in the clumpy ejecta. The metallic grain masses thus increase by $\sim$~3~orders of magnitude compared to the homogeneous case. 
%This has a significant impact on the overall condensation picture. The average grain sizes for the first category of dust components vary at most by a factor of 10 from standard to special case. The grain sizes for forsterite however is found to increase from 60~\AA to 3200~\AA, and for metallic clusters by three orders of magnitude in the clumps. 
%Moreover the study of the clumpy model indicates the presence of a small of mass of very large grains ($\sim \mu$) and a large mass of moderate ($\sim$ 500 \AA) in the ejecta, which is unlike the standard case. This might be an interesting case study to access the further processing of dust in the shocks.  

The mass distribution as a function of grain size is shown in Figure \ref{fig6} (bottom-right panel), and the masses for the various dust components are summarised in Table \ref{tab7}, along with the $a_{peak}$ radii that characterise each grain components. The distribution is skewed towards large grain sizes comprised between 100~\AA\ and 1 \mic, which contain most of the mass of forsterite, pure iron and silicon, alumina, and carbon grains. As seen from Table \ref{tab7}, the alumina and carbon mass are similar to those for the homogeneous case as their final masses are controlled by the initial abundances of Al in the ejecta and the C/O ratio of zone 5, respectively. Of interest is the strong contribution of $\sim 0.1$ \mic\ grains of pure iron and silicon to the total dust mass, while these grains have a minor contribution to the size and mass distributions for the homogeneous case. The clumpy ejecta of the 19~\Ms\  progenitor forms a total dust mass of $0.14$~\Ms, which is greater than the dust mass derived for the homogeneous case  by a factor of four. 

%--------------- Fig 7 ----------------
\begin{figure}
 %\centering
\includegraphics[width=\columnwidth]{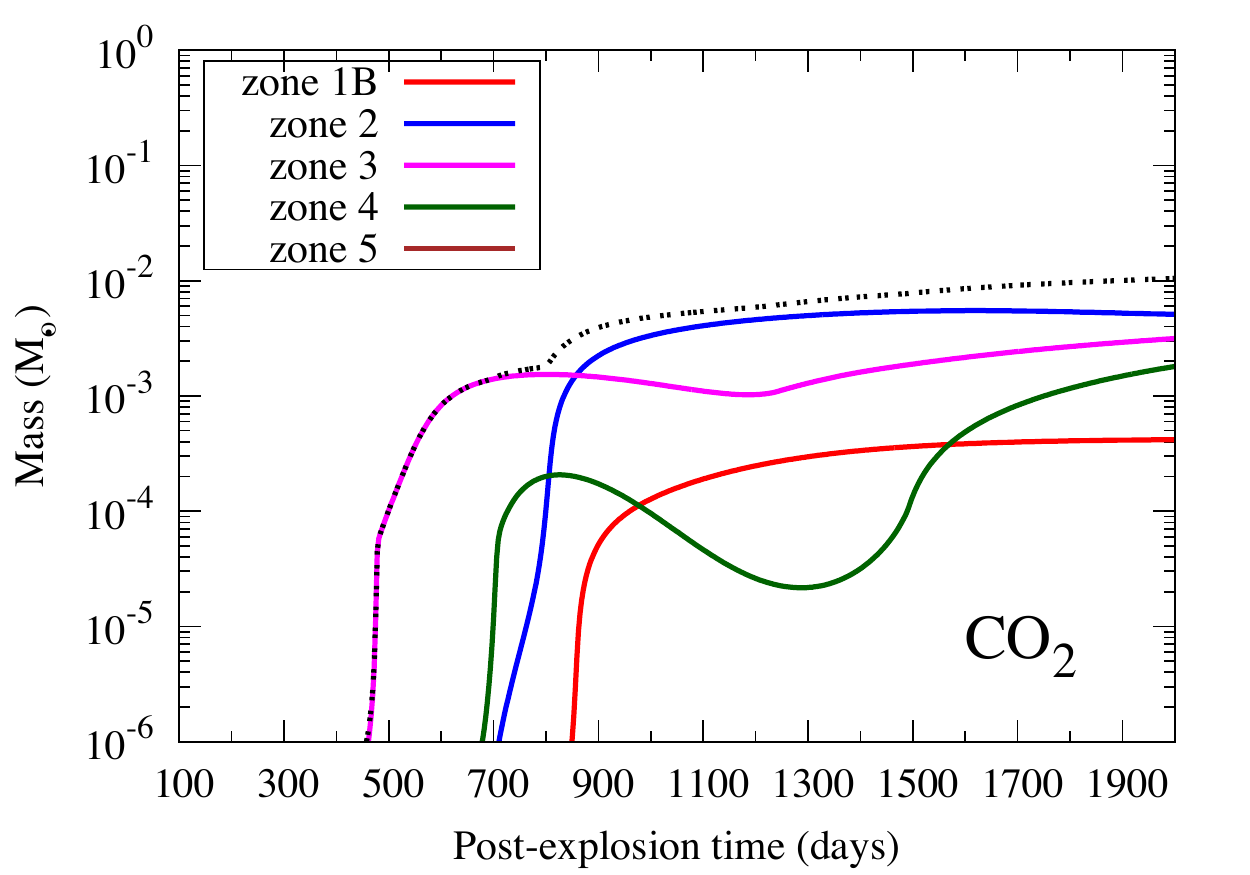}
 \caption{Mass of CO$_2$ formed in the clumpy ejecta of the 19~\Ms\ progenitor as a function of post-explosion time and ejecta zone.}
 \label{fig7}
\end{figure}
%-------------------------------------

The high gas densities of the clumpy model enhance the efficiency of the gas-phase chemistry and boost the formation of specific molecules in the ejecta. This is the case for CO$_2$, as shown in Figure \ref{fig7}, where the mass as a function of post-explosion time and zoning is shown. The molecule forms at day 500 in zone 3, and grows in mass from a second formation event in zone 2 starting at day 800. At day 2000, the total CO$_2$ mass reaches $ \sim 1 \times 10^{-2}$~\Ms. The molecules N$_2$ and NO also form with a mass of $7.5 \times 10^{-3}$~\Ms\ and $1.1 \times 10^{-5}$~\Ms, respectively, at day 2000. In the homogeneous case, CO$_2$ forms with a low mass of $3\times 10^{-4}$~\Ms. A large reservoir of CO molecules is produced in the clumpy case, where the CO mass reaches $\sim 0.06$ \Ms\, $0.1$ \Ms, and $0.3$ \Ms\ at day 2000 in zones 3, 2, and 4, respectively. A smaller mass of CO ($6\times 10^{-3}$ \Ms) is produced in the carbon-rich zone 5, where carbon dust condenses. Thus, CO cannot be considered as a carbon dust tracer in SN ejecta, as already pointed out by SC13. The large CO reservoir in zones 2 and 3 provides an effective formation channel for CO$_2$ via CO recombination. Carbon dioxide is thus a tracer of clumpiness in the O-rich core of the 19 \Ms\ ejecta. 

%--------------- Fig 8 ----------------
\begin{figure*}
 \centering
\includegraphics[width=19cm]{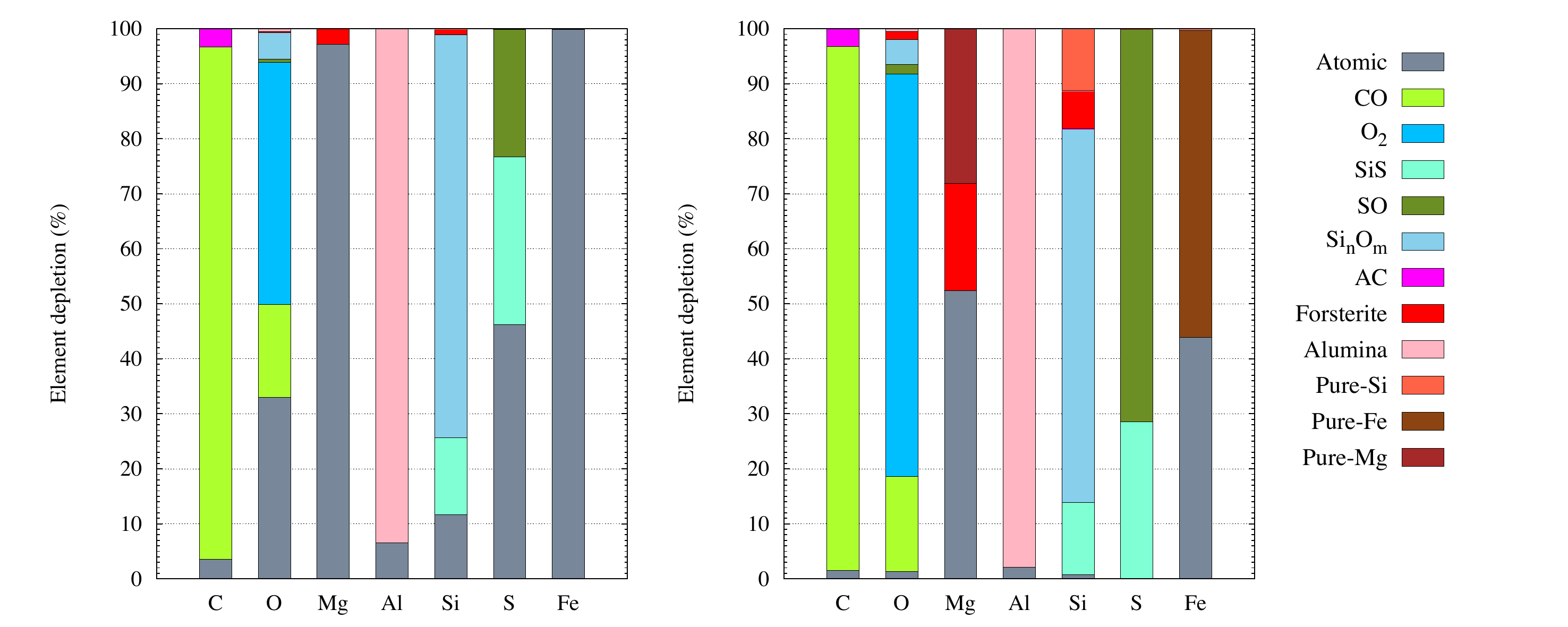}
\qquad
 \caption{Elemental depletion for the homogeneous (left) and the clumpy (right) ejecta of the 19~\Ms\ progenitor at day 2000 post-explosion.}
 \label{fig8}
\end{figure*}
%-------------------------------------

\subsubsection{Elemental depletion}

We calculate the depletion of elements in molecules, dust clusters and grains to assess the effect of ejecta clumps on the depletion fraction. Results for both homogeneous and clumpy SN ejecta with the 19 \Ms\ stellar progenitor are shown in Figure \ref{fig8} for the various elements of interest. 

In the homogeneous case, the fraction of elements staying in atomic form is high for oxygen and sulphur, and almost 100 \% for both magnesium and iron. Carbon is essentially depleted in CO for both homogeneous and clumpy ejecta, with a few \% going to amorphous carbon. The oxygen depletion changes drastically in the clumpy ejecta with $\sim 70$ \% of oxygen in the form of O$_2$, and a larger depletion in SO and forsterite. While more than 96 \% of magnesium is atomic for the homogeneous ejecta, almost 50 \% of magnesium is trapped in forsterite and pure Mg when the ejecta is clumpy. For both cases, aluminium is heavily depleted in alumina, with a fraction exceeding 93 \%, and the rest left in atomic form. 
Sulphur is not depleted in metal sulphides for both cases as the amount of FeS formed in both ejecta is small (see Table \ref{tab7}). Sulphur is depleted in the molecules SO and SiS, and the depletion becomes total for the clumpy case. Finally, clumpiness has a strong effect on iron. In the homogeneous ejecta, almost all iron is in atomic form since the mass of formed FeS is very small. However, in the clumpy case, 56 \% of iron is in large grains of pure iron, as seen in Figure \ref{fig6}. 

For both cases, we see that a high fraction of Si atoms is trapped in clusters which enter the formation process of silica, SiO$_2$. These clusters labelled, Si$_x$O$_y$, will not all be included in the final silica mass, as their growth process is controlled by the amount of available SiO. As mentioned in \S\ \ref{sec3}, we have not studied the nucleation and condensation of silica in this study, but an assessment of the silica mass based on the SiO mass at day 300 or the assumption that all Si$_x$O$_y$ clusters turn into silica leads to values ranging between $10^{-5}$ \Ms\ and $10^{-2}$ \Ms, respectively, at day 2000. 

More generally, we see that a clumpy ejecta depletes almost all elements in molecules, dust clusters and grains, except for magnesium and iron, which retain some high mass fraction in atomic form. However, in contrast with the homogeneous ejecta, clumpiness also favours the depletion of $\sim$ 46 \% of the magnesium mass in pure magnesium and silicate grains, whereas $\sim$ 56 \% of the iron mass is depleted in pure iron grains. 

\section{Summary and discussion}

\label{sec5}
We have presented an exhaustive model of dust synthesis in the homogeneous and clumpy ejecta of Type II-P SNe, where the gas-phase chemistry, including the formation of dust clusters (i.e.,~nucleation phase), is coupled to the coagulation and coalescence of these clusters into dust grains (i.e., condensation phase). Our findings are summarised below. 

\begin{itemize}
\item{As soon as dust clusters form from the gas phase, the coalescence and coagulation of these clusters is very efficient at growing dust grains with an appreciable size for all models. The effective condensation of grains confirms the crucial role of the nucleation phase as a bottleneck to dust production, as already highlighted by SC13. The nucleation of clusters indeed controls the chemical type and the final mass of the dust produced in SN ejecta.} 
\item{All SN ejecta form three prevalent populations of dust: silicates, carbon and alumina. Homogeneous ejecta form medium-size grains with size distributions peaking between $50$ \AA\ and $100$ \AA\ for these three dust components. However, the prevalence of a dust type depends on the elemental composition of the stellar progenitor before explosion, hence the progenitor mass. A low-mass progenitor predominantly forms carbon dust while a high-mass progenitor forms silicates and alumina. }
\item{The grain size distributions for all cases are different from a power-law distribution or the MRN power-law size distribution representative of interstellar dust. Specifically, all our derived size distributions are skewed towards large grains, with several distribution peaks that correspond to dust condensation events at different epochs and positions in the ejecta zones.} 
\item{A clumpy ejecta and/or a low \Ni\ mass favour the formation of dust clusters and grains at early post-explosion time in the various zones. These grains have thus more time to grow in the nebular phase, and lead to the formation of large grains with size over $0.1$ \mic. Furthermore, clumps boost the formation of pure metallic grains, including iron and silicon dust in the innermost He-core zones, and magnesium grains in the oxygen-rich zones. These metallic grains have individual masses comparable to that of alumina, and when summed over the three dust populations, their total mass is comparable to that of silicates. Iron sulphide grains represent a minor dust component for all ejecta cases. }
\end{itemize}

%--------------- Fig 9 ----------------
\begin{figure*}
 \centering
\includegraphics[width=19cm]{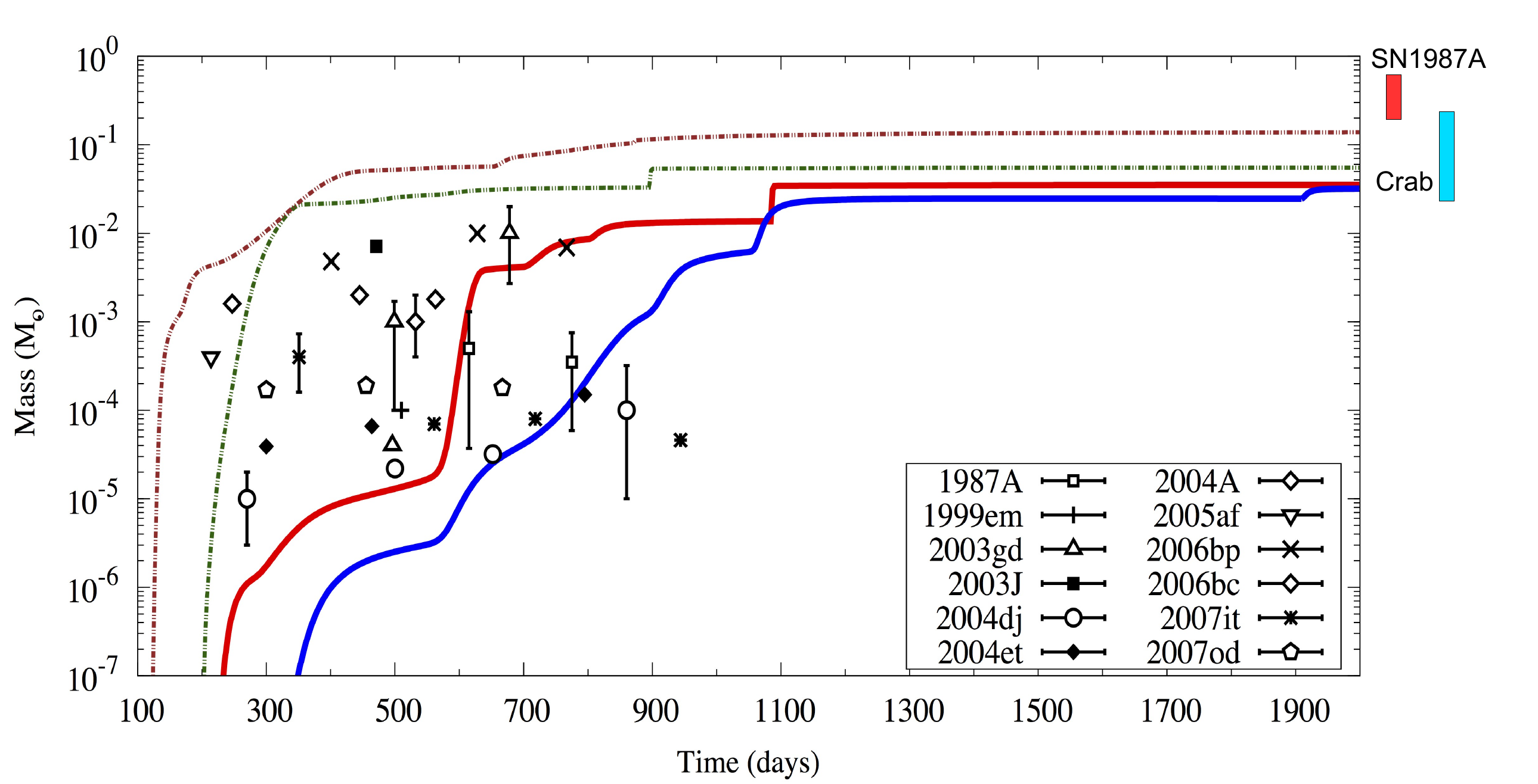}
\qquad
 \caption{Total modelled dust masses versus post-explosion time for the various SN cases considered in this study and dust masses derived from mid-IR data for a sample of Type II-P SNe - SN1987A (\cite{wood93, erc07}), SN1999em (\cite{elm03}), SN2003gd (\cite{sug06,meik07}), SN2003J (\cite{sza13}), SN2004dj (\cite{sza11, mei11}), SN2004et (\cite{ko09}), SN2004A (\cite{sza13}), SN2005af (\cite{kot08}), SN2006bp (\cite{gal12, sza13}), SN2007it (\cite{and11}), SN2007od (\cite{ins11, and10}). The bars associated with the data points indicate the range of mass values derived from various authors or models for a specific SN. Full blue and red lines:  19 \Ms\ and 15 \Ms\ homogeneous ejecta, respectively. Green dotted line: low \Ni\ mass, 15 \Ms\ homogeneous ejecta. Brown dotted line: 19 \Ms\ clumpy ejecta. The dust masses derived for the young remnant SN1987A and the Crab Nebula are also shown (SN1987A - \cite{mat11,ind14}, The Crab - \cite{gom12, tem13}).}
 \label{fig9}
\end{figure*}
%-------------------------------------

The modelling of the IR emission flux due to warm dust in Type II SN ejecta usually involves the use of a MRN power-law size distribution with index $\alpha \sim -3.5$ for dust grains. Two types of dust are assumed, either carbon or silicates, or a mixture of both, and dust masses are derived (e.g., \cite{sug06, erc07, ko09}). As our results illustrate, the final size distributions obtained for dust grains that condense in homogenous and clumpy ejecta do not follow such a power-law distribution. Furthermore, the dust chemical composition is much more complex than those usually assumed, is extremely time-dependent, and include several dust components with grain size distributions that reflect the He-core mass position and time at which dust forms. So fitting the dust contribution to the spectral energy distribution at different times requires the use of different dust chemical compositions and size distributions. This differs strongly from assuming a simple MRN distribution for one or two dust components. 

Our results show that the synthesis of dust highly depends on two parameters, the mass of \Ni\ produced in the explosion and the ejecta gas density. A low \Ni\ mass favours the synthesis of dust at early post-explosion time in the various ejecta zones predominantly forming silicates, metal oxide, and carbon. For example, in SN 2003gd, which has a low-mass progenitor and a low \Ni~ mass, a mid-IR excess, along with asymmetric blue-shifted emission lines and an increase in optical extinction, was observed as evidence for dust formation in the ejecta, as early as day 250 (\cite{hen05, sug06}).

The dependence on gas density is even stronger and is well illustrated by our clumpy ejecta case. A clumpy ejecta favours the dust formation in all ejecta zones at early times, and result in several population of large grains a few years after explosion. Indeed, dust forms in the dense ejecta zones, and the early-formed grains have then time to grow to fairly large grains in the ejecta. 
A large fraction of these large grains will survive the non-thermal sputtering induced by the reverse shock, sheltered in the dense ejecta clumps, and the thermal sputtering in the hot, inter-clump medium once the clumps are disrupted during the SNR phase (Biscaro \& Cherchneff, in preparation). Some graphite and silicate pre-solar grains found in meteorites have a SN origin (\cite{zin07, hop10}). According to the present results, they may be identified as the largest grains of silicates and carbon that form in the dense ejecta clumps of Type II-P SNe. However, our results cannot explain the elemental mixing between ejecta zones as derived from the isotopic ratio analysis of these pre-solar grains, because our models use stratified ejecta with no leakage between zones. Clues on elemental mixing from different ejecta zones are provided by 3-D simulations of SN explosion from which the chemical composition of clumps are derived a few hours after outburst (e.g., \cite{ham10}), and further studies will consider these outputs as initial conditions for the physico-chemical model of clumpy SN ejecta. 

Most important is the fact that grains respond differently to gas density enhancement. Silicate dust production is extremely dependent on gas density because the nucleation phase of this type of grains is characterised by complex chemical pathways, which are density-dependent. The nucleation phase thus controls the final amount of silicate dust mass that forms in the SN explosion, as shown by SC13. Other tracers of density increase in the ejecta of Type~II-P~SNe are pure metal dust, such as silicon, magnesium, and iron grains, and new molecules that only form with high abundance and mass in the clumpy ejecta, i.e., CO$_2$. 

Conversely, the production of alumina and carbon is not too responsive to density increase and is limited by the availability of atomic carbon and aluminium in the ejecta zones where these specific dust grains form. For example, according to Figure \ref{fig8} and for all SN progenitor masses, atomic carbon is essentially depleted in CO molecules, which primarily form in the oxygen-rich zone labelled 4, and in carbon dust produced in the outermost, C-rich zone 5. The efficiency at forming carbon dust depends on the C/O ratio of this ejecta zone. For the 15 \Ms\ progenitor, the C/O ratio of zone 5 is high ($\sim 21$), and thus 78 \% of the carbon mass yield gets into carbon dust while $\sim$ 20 \% stay in the form of carbon chains in the gas phase. For the 19 \Ms\ progenitor, the C/O is small ($\sim 4$), and only 58 \% of the initial carbon mass yield gets into grains for both homogeneous and clumpy ejecta. The final mass of carbon dust formed in Type II-P SN ejecta is thus limited by the carbon mass yield of the outermost, carbon-rich, ejecta zone, and not by the total carbon yield of the ejected material, as it is often assumed (e.g., Matsuura et al. 2011, 2014). Although carbon is one important component of SN dust~$\sim$~5 years post-outburst, silicates, alumina, and pure metals are also important dust components, especially when ejecta clumpiness is taken into account. Actually, a satisfactory fit of the Herschel data on SN1987A requires a population of large ($a=0.5$ \mic) iron grains (\cite{mat11}). Hence, the analysis of the submm flux emitted by cool, thermal dust in SN remnants must consider these various dust components as a whole to properly assess the mass of dust formed in the ejecta. 

We conclude that the synthesis of dust in Type II-P is a multi-parameter-dependent process. To illustrate this point, our total modelled dust masses for various SN progenitors, the dust masses derived from mid-IR observation of several Type II-P SNe, and the dust masses assessed from submm data of SN remnants are plotted in Figure \ref{fig9}. The dust masses derived before day 1000  from mid-IR data span a high value range comprised between $10^{-6}$ \Ms\ and $10^{-2}$ \Ms. This large spread in dust mass reflects the difference in SN progenitor mass, the mass of \Ni, the likely presence of ejecta clumps, and the chemical composition of the dust that forms. Our modelled dust mass values for various progenitors, \Ni\ mass, and clumpy/non clumpy ejecta models well reproduce this large spread, and point to the fact that dust synthesis in SNe depends on several parameters, and cannot be described by assuming a simple dust composition, size distribution, and dust temperature. As already pointed out by SC13, our results indicate a gradual increase of dust production and growth over a time span of $\sim 3-5$ years after explosion, to reach dust masses in the range $0.03-0.2$~\Ms. This range agrees well with the latest submm dust masses derived from Herschel and ALMA data for SN1987A and other SN remnants. The trend of a gradual growth of dust grains over $\sim 3-5$ years is in contrast with recent radiative transfer study of SN1987A by Wesson et al. (2015), who conclude the major growth of dust grains has taken place between day 1200 and day 9200, and that the grains are very large ($a\sim 3-5$ \mic). They argue that growth by atom accretion on the grain surface is not effective enough to form such large grains and propose coagulation instead. However, coagulation is usually characterised by larger time scales than accretion and is effective at high gas densities, as shown in this study. Therefore, the mechanism triggering dust growth at very late post-explosion time and low gas density has yet to be identified. The dust mass values derived in the present study are still well below the 1~\Ms\ required to account for the observed dust mass in J1148+5251 at z=6.4. However primeval SNe are expected to be more massive than their local counterparts considered in this study (\cite{hir14}), and may produce larger dust masses (\cite{noz03, sch04, cher10}). 

The presolar grains of SN origin found in meteorites may have formed in massive and dense Type~II-P~SNe akin to our clumpy 19 \Ms\ surrogate model for SN1987A, as these large dust grains may survive the remnant phase sheltered in their dense, clumpy cradle. Out of the twelve Type II-P SNe plotted in Figure \ref{fig9}, eight~SNe have masses of warm dust detected before 1100 days that agree with our homogeneous SN models, and for which rather small grains are synthesised. These grains may not survive the remnant phase. This study points to the fact that while some dense Type~II-P SNe may contribute to providing dust to the Interstellar Medium and the solar system, many Type~II-P~SNe will probably be modest contributors to the dust budget of local galaxies.

\begin{acknowledgements}
The authors thank the anonymous referee for useful comments that helped improving the manuscript, John Plane for providing the initial version of the condensation code, and Rubina Kotak and Stefan Bromley for stimulating discussions. A.S. acknowledges support from the Swiss National Science Foundation through the subside 20GN21?128950 linked to the CoDustMas network of the European Science Foundation Eurogenesis programme.
\end{acknowledgements}

\appendix
\section{}
% -------------------------- Table 1 Type IIb initial composition 15-19 Msun -------------------
\begin{table*}
\caption{Gas mean molecular weight $\mu_{gas}$, C/O ratio, and initial elemental mass yields as a function of ejecta zone for the 15 and 19~\Ms\ stellar progenitors (SC13). The total yield is the elemental mass yield for the total ejected material.}
\label{tab1} 
\centering
\begin{tabular}{l l l l l   l l l l l l l l l}
\hline \hline
\multicolumn{1}{c}{Zone} & \multicolumn{1}{c}{$\mu_{gas}$}& \multicolumn{1}{c}{C/O}& \multicolumn{1}{c}{He} & \multicolumn{1}{c}{C}& \multicolumn{1}{c}{O} &\multicolumn{1}{c}{Ne} & \multicolumn{1}{c}{Mg}&\multicolumn{1}{c}{Al}& \multicolumn{1}{c}{Si}& \multicolumn{1}{c}{S }&\multicolumn{1}{c}{Ar}& \multicolumn{1}{c}{Fe} & \multicolumn{1}{c}{Ni} \\
\hline
\multicolumn{14}{c}{15~\Ms\ progenitor}\\
\hline
1A & 35.5& 5.9(-2) & 0 & 1.5(-7) & 3.3(-6) & 0 & 1.4(-5) & 2.0(-5) & 3.2(-2) & 2.0(-2) & 4.0(-3) & 1.7(-2) & 2.8(-4) \\
1B  & 20.9& 2.1(-3) & 0 & 6.9(-6) & 4.4(-2) & 1.0(-5) & 3.9(-4) & 5.0(-5) & 3.1(-2) & 1.3(-2) & 7.4(-4) & 1.3(-4) & 1.3(-7) \\
2  & 17.2 & 5.5(-3) & 0 & 9.3(-4) & 0.23 & 1.5(-2) & 1.6(-2) & 2.1(-3) & 2.1(-2) & 2.5(-3) & 4.1(-5) & 2.3(-5) & 0 \\
3 & 17.1 & 1.6(-2) & 0 & 2.8(-3) & 0.24 & 7.8(-2) & 1.8(-2) & 1.9(-3) & 1.8(-3) & 6.8(-5) & 1.7(-5) & 2.3(-5)  & 0 \\
4A &15.0 & 0.37 & 6.1(-6) & 4.0(-2) & 0.15 & 3.0(-3) & 1.6(-4) & 2.3(-4) & 7.1(-5) & 3.5(-5) & 9.6(-6) & 2.2(-5)  & 0 \\
4B  & 10.7 & 0.74 & 3.1(-2) & 6.2(-2) & 0.11 & 1.4(-2) & 7.1(-4) & 1.9(-5) & 1.1(-4) & 4.4(-5) & 8.6(-6) & 4.0(-5)  & 0 \\
5  & 4.1 & 21.3 & 0.71 & 2.7(-2) & 1.7(-3) & 1.2(-3) & 3.9(-4) & 5.3(-5) & 4.8(-4) & 2.9(-4) & 1.2(-5) & 8.4(-4) &  0 \\
6  & 4.1 & 1.2 & 0.34 & 9.1(-5) & 9.6(-5) & 5.5(-4) & 1.8(-4) & 2.4(-5) & 2.3(-4) & 1.4(-4) & 5.3(-6) & 4.1(-4) &  0 \\
\hline
\multicolumn{3}{l}{Total Yield} & 1.1& 0.13& 0.78  &0.11 & 3.6(-2) & 4.4(-3) & 8.7(-2) & 3.6(-2) &4.8(-3) & 1.8(-2)&  2.8(-4) \\
\hline
\multicolumn{14}{c}{19~\Ms\ progenitor}\\
\hline
1A & 35.3 & 0.16 & 1.4(-6) & 8.1(-8) & 6.9(-7) & 0 & 1.7(-5) & 2.5(-5) & 3.8(-2) & 2.3(-2) & 4.5(-3) & 2.5(-2) &  3.3(-4) \\
1B  & 22.5& 1.3(-3) & 0 & 1.2(-4) & 0.12 & 1.1(-4) & 8.8(-4) & 2.2(-4) & 9.9(-2) & 5.6(-2) & 1.5(-2) & 3.1(-3)  & 6.2(-6) \\
2  & 16.9 & 6.5(-2) & 0 & 5.9(-2) & 1.2 & 0.3 & 8.4 (-2) & 9.1(-3) & 1.5(-2) & 1.3(-3) & 1.2(-4) & 7.5(-4) &  0 \\
3  & 15.1 & 0.4 & 0 & 2.9(-2) & 0.11 & 2.8(-3) & 2.0(-3) & 1.6(-5) & 1.1(-4) & 3.2(-5) & 8.9(-6) & 7.8(-5)  & 0 \\ 
4 & 10.3 & 0.6 & 7.7(-2) & 0.13 & 0.26 & 1.3(-2) & 5.4(-3) & 5.3(-5) & 3.8(-4) & 1.4(-4) & 3.5(-5) & 3.8(-4)  & 0 \\
5 & 4.1 & 3.9 & 0.74 & 1.3(-2) & 4.3(-3) & 1.1(-2) & 5.0(-4) & 6.0(-5) & 5.5(-4) & 3.2(-4) & 7.0(-5) & 9.8(-4)  & 0 \\
6 & 4.1 & 1.8 & 0.35 & 1.7(-4) & 9.2(-5) & 5.7(-4) & 2.3(-4) & 3.2(-5) & 2.6(-4) & 1.5(-4) & 3.3(-5) & 4.6(-4)  & 0 \\
\hline
\multicolumn{3}{l}{Total Yield} & 1.2& 0.23 &1.69  &0.33 &9.3(-2) &9.5(-3) &0.15 & 8.1(-2)&2.0(-2) & 3.7(-2)&  3.4(-4) \\
\hline 
\end{tabular}
\tablefoot{Mass yields are in~\Ms\ and the gas mean molecular weight $\mu_{gas}$ is in g \cmc. Parameters for the SN ejecta with a progenitor mass of $=15$~\Ms: explosion energy~E$_{kin}=1\times 10^{51}$~erg; He-core mass~$=4.14$~~\Ms; ejecta velocity $=2000$ \kms; and \gray~Optical depth $\tau_{\gamma} = 17.5$. Parameters for the SN ejecta with a progenitor mass of $=19$~\Ms: explosion energy~E$_{kin}=1\times 10^{51}$~erg; He-core mass~$=5.62$~\Ms; ejecta velocity $=2000$ \kms; and \gray~Optical depth $\tau_{\gamma} = 23$. Nitrogen, N, is only present in Zone 6 with a yield of $4.3\times 10^{-3}$ for the 15~\Ms\ progenitor and $4.5\times 10^{-3}$ for the 19~\Ms\ progenitor.}\\

\end{table*}
%-------------------------------------------------
%\clearpage

%\section{The physical gas parameters of SN ejecta with a 15 \Ms\ and 19 \Ms\ stellar progenitors. }

%---------------------------- Table 2 Gas parameters in  the ejecta --------------------------
%\begin{comment} 
\begin{table*}
\caption{Ejecta temperature T$_{gas}$ and number density n$_{gas}$ for the homogeneous, stratified SN ejecta with 15 and 19~\Ms\ stellar progenitors as a function of post-explosion time and ejecta zones (SC13). The mass coordinates of each zone in the He-core are listed. }
\label{tab2}   
\centering
\begin{tabular}{ c  l l  l l  l l  l l  l l  l l  l l }
\hline\hline
\multicolumn{15}{c}{15~\Ms\ progenitor}\\
\hline
Zones & \multicolumn{2}{c}{1A (1.79-1.88)} & \multicolumn{2}{c}{1B (1.88-1.98)} & \multicolumn{2}{c}{2 (1.98-2.27)} & \multicolumn{2}{c}{3 (2.27-2.62)} & \multicolumn{2}{c}{4A (2.62-2.81)} & \multicolumn{2}{c}{4B (2.81-3.04)} &  \multicolumn{2}{c}{5 (3.04-3.79)} \\
%}
\hline
Day & T & n$_{gas}$ & T & n$_{gas}$ & T & n$_{gas}$ & T & n$_{gas}$& T & n$_{gas}$ & T & n$_{gas}$ & T & n$_{gas}$  \\
\hline
100 & 12000 & 1.8(11) & 11600 & 3.1(11) & 10400 & 3.7(11) & 8779 & 3.8(11) & 7980 & 4.3(11) & 7580  & 6.1(11) & 6490 & 1.6(12) \\
300 & 3006 & 6.7(9) & 2906 & 1.1(10) & 2605 & 1.4(10)  & 2199 & 1.4(10) & 1998 & 1.6(10) & 1899  & 2.3(10) & 1626 & 5.9(10)  \\
600 & 1255 & 8.3(8) & 1213 & 1.4(9) & 1088 & 1.7(9) & 918 & 1.8(9) & 835 & 2.0(9) & 793 & 2.8(9) & 679 & 7.4(9) \\
900 & 753 & 2.5(8) & 728 & 4.3(8) & 653 & 5.1(8)&  551 & 5.2(8) & 501 & 5.9(8) & 476  & 8.4(8) & 407 & 2.2(9)  \\
1200 & 524 & 1.0(8) & 507 & 1.8(8) & 454 & 2.1(8)&  383 & 2.2(8) & 349 & 2.5(8) & 331 & 3.5(8) & 283 & 9.3(8)  \\
1500 & 396 & 5.3(7) & 382 & 9.2(7) & 343 & 1.1(8)&  289 & 1.1(8) & 263 & 1.3(8) & 250 & 1.8(8) & 214 & 4.7(8)  \\
2000 & 275 & 2.3(7) & 266 & 3.9(7) & 239 & 4.6(7) &  201 & 4.8(7) & 183 & 5.4(7) & 174 & 7.6(7) & 149 & 2.0(8)  \\
\hline 
\multicolumn{15}{c}{19~\Ms\ progenitor}\\
\hline
Zones & \multicolumn{2}{c}{1A (1.77-1.88)} & \multicolumn{2}{c}{1B (1.88-2.18)} & \multicolumn{2}{c}{2 (2.18-3.86)} & \multicolumn{2}{c}{3 (3.86-4.00)} & \multicolumn{2}{c}{4 (4.00-4.49)} & \multicolumn{2}{c}{5 (4.49-5.26)} &\multicolumn{2}{c}{6 (5.26-5.62)}  \\
\hline
Day& T & n$_{gas}$ & T & n$_{gas}$ & T & n$_{gas}$ & T & n$_{gas}$& T & n$_{gas}$ & T & n$_{gas}$ & T & n$_{gas}$  \\
\hline
100 & 12400 & 7.2(10) & 12000 & 1.2(11) & 9980 & 1.5(11) & 7190 & 1.7(11) & 6390 & 2.5(11) & 6000  & 6.3(11) & 5900 & 6.4(11)  \\
300 & 3106 & 2.7(9) & 3006 & 4.3(9) & 2500 & 5.7(9)  & 1801 & 6.4(9) & 1601 & 9.3(9) & 1503  & 2.3(10) &  1478 & 2.4(10)  \\
600 & 1297 & 3.3(8) & 1255 & 5.4(8) & 1044 & 7.1(8) & 752 & 8.0(8) & 668 & 1.2(9) & 628 & 2.9(9) & 617 & 3.0(9)  \\
900 & 778 & 9.8(7) & 753 & 1.6(8) & 626 & 2.1(8)&  451 & 2.4(8) & 401 & 3.5(8) & 377  & 8.7(8) &370 & 8.8(8)   \\
1200 & 542 & 4.1(7) & 524 & 6.7(7) & 436 & 8.9(7)&  314 & 1.0(8) & 279 & 1.5(8) & 262 & 3.7(8) & 258 & 3.7(8) \\
1500 & 409 & 2.1(7) & 396 & 3.4(7) & 329 & 4.6(7) &  237 & 5.1(7) & 211 & 7.5(7) & 198 & 1.9(8) & 195 & 1.9(8)   \\
2000 & 285 & 9.0(6) & 275 & 1.5(7) & 229 & 1.9(7) &  165 & 2.2(7) & 147 & 3.2(7) & 138 & 7.9(7) & 135 & 8.0(7)  \\
\hline
\end{tabular}
\tablefoot{Zone mass coordinates are in~\Ms, T$_{gas}$ in Kelvin and n$_{gas}$ in \cmc.} \\
\end{table*}
% -----
%-------------------------Table ------------------------
\begin{table*}
\caption{Grain size distribution function $f(a)$ normalised to $N_{tot}(a)$ as a function of size $a$ at day 2000 for all the dust components of the standard 15 \Ms\ case.}
\label{taba3}
\centering
\begin{tabular}{r c r c r c r c}
\hline \hline
$a$ (\AA) & $f(a)$ & $a$ (\AA) & $f(a)$ & $a$ (\AA) & $f(a)$ & $a$ (\AA) & $f(a)$ \\
\hline
\multicolumn{2}{c}{Forsterite} & \multicolumn{2}{c}{Alumina} & \multicolumn{2}{c}{Carbon} & \multicolumn{2}{c}{Pure-Si} \\
\hline
3.33	&	5.76(-01)	&	3.45	&	6.20(-01)	&	3.93	&	2.43(-02)	&	2.46	&	6.23(-01)	\\
4.05	&	1.29(-01)	&	4.19	&	1.38(-01)	&	4.78	&	3.35(-05)	&	2.99	&	1.42(-01)	\\
4.93	&	8.55(-02)	&	5.10	&	8.84(-02)	&	5.81	&	1.01(-05)	&	3.63	&	9.13(-02)	\\
6.00	&	5.44(-02)	&	6.20	&	5.10(-02)	&	7.07	&	5.48(-07)	&	4.42	&	5.21(-02)	\\
7.30	&	4.14(-02)	&	7.55	&	3.26(-02)	&	8.60	&	5.55(-08)	&	5.38	&	3.17(-02)	\\
8.88	&	3.51(-02)	&	9.18	&	2.25(-02)	&	10.46	&	3.41(-09)	&	6.54	&	1.98(-02)	\\
10.80	&	2.95(-02)	&	11.17	&	1.63(-02)	&	12.72	&	8.30(-10)	&	7.96	&	1.27(-02)	\\
13.14	&	2.14(-02)	&	13.59	&	1.13(-02)	&	15.48	&	6.63(-08)	&	9.68	&	8.24(-03)	\\
15.98	&	1.22(-02)	&	16.53	&	6.91(-03)	&	18.83	&	2.33(-06)	&	11.77	&	5.41(-03)	\\
19.44	&	6.00(-03)	&	20.10	&	3.81(-03)	&	22.90	&	3.78(-05)	&	14.32	&	3.56(-03)	\\
23.64	&	3.05(-03)	&	24.46	&	2.13(-03)	&	27.86	&	3.33(-04)	&	17.42	&	2.33(-03)	\\
28.76	&	1.64(-03)	&	29.75	&	1.21(-03)	&	33.89	&	1.82(-03)	&	21.19	&	1.50(-03)	\\
34.99	&	9.89(-04)	&	36.19	&	6.49(-04)	&	41.23	&	6.83(-03)	&	25.78	&	9.49(-04)	\\
42.56	&	8.15(-04)	&	44.02	&	4.16(-04)	&	50.15	&	1.92(-02)	&	31.36	&	6.45(-04)	\\
51.77	&	7.75(-04)	&	53.55	&	4.80(-04)	&	61.00	&	4.30(-02)	&	38.14	&	5.82(-04)	\\
62.97	&	6.36(-04)	&	65.14	&	6.95(-04)	&	74.21	&	8.09(-02)	&	46.40	&	6.97(-04)	\\
76.60	&	4.07(-04)	&	79.24	&	9.08(-04)	&	90.27	&	1.31(-01)	&	56.44	&	8.65(-04)	\\
93.18	&	2.49(-04)	&	96.39	&	9.78(-04)	&	109.81	&	1.81(-01)	&	68.66	&	9.38(-04)	\\
113.35	&	2.12(-04)	&	117.25	&	8.16(-04)	&	133.57	&	2.04(-01)	&	83.52	&	7.99(-04)	\\
137.88	&	2.15(-04)	&	142.62	&	4.98(-04)	&	162.48	&	1.73(-01)	&	101.59	&	4.76(-04)	\\
167.73	&	1.94(-04)	&	173.49	&	2.16(-04)	&	197.65	&	9.75(-02)	&	123.58	&	1.77(-04)	\\
204.03	&	1.36(-04)	&	211.04	&	6.33(-05)	&	240.43	&	3.17(-02)	&	150.33	&	3.73(-05)	\\
248.19	&	6.58(-05)	&	256.72	&	1.15(-05)	&	292.47	&	5.29(-03)	&	182.87	&	4.27(-06)	\\
301.91	&	1.92(-05)	&	312.29	&	1.21(-06)	&	355.77	&	4.27(-04)	&	222.45	&	2.75(-07)	\\
367.26	&	3.13(-06)	&	379.88	&	7.28(-08)	&	432.77	&	1.71(-05)	&	270.60	&	1.10(-08)	\\
446.74	&	2.77(-07)	&	462.10	&	2.68(-09)	&	526.44	&	3.76(-07)	&	329.16	&	3.06(-10)	\\
543.44	&	1.41(-08)	&	562.11	&	6.30(-11)	&	640.39	&	5.03(-09)	&	400.41	&	6.04(-12)	\\
661.06	&	4.41(-10)	&	683.78	&	9.99(-13)	&	778.99	&	4.36(-11)	&	487.07	&	8.34(-14)	\\
804.14	&	9.02(-12)	&	831.78	&	1.09(-14)	&	947.60	&	2.54(-13)	&	592.49	&	8.21(-16)	\\
978.19	&	1.25(-13)	&	1011.80	&	8.47(-17)	&	1152.70	&	1.02(-15)	&	720.73	&	5.85(-18)	\\
1189.90	&	1.21(-15)	&	1230.80	&	4.69(-19)	&	1402.20	&	2.87(-18)	&	876.73	&	3.05(-20)	\\
1447.50	&	8.32(-18)	&	1497.20	&	1.89(-21)	&	1705.70	&	5.76(-21)	&	1066.50	&	1.17(-22)	\\
1760.70	&	4.10(-20)	&	1821.30	&	5.56(-24)	&	2074.90	&	8.36(-24)	&	1297.30	&	3.36(-25)	\\
2141.80	&	1.48(-22)	&	2215.40	&	1.21(-26)	&	2523.90	&	8.85(-27)	&	1578.10	&	7.20(-28)	\\
\hline
\end{tabular}
\end{table*}
%----------
\begin{table*}
\caption{Grain size distribution function $f(a)$ normalised to $N_{tot}(a)$ as a function of size $a$ at day 2000 for all the dust components of the standard 15 \Ms\ case.}
\label{taba3}
\centering
\begin{tabular}{r c r c r c r c}
\hline \hline

$a$ (\AA) & $f(a)$ & $a$ (\AA) & $f(a)$ & $a$ (\AA) & $f(a)$ & $a$ (\AA) & $f(a)$ \\
\hline
\multicolumn{2}{c}{Pure-Mg} & \multicolumn{2}{c}{Pure-Fe} & \multicolumn{2}{c}{FeS} & \multicolumn{2}{c}{SiC} \\
\hline
2.28	&	6.05(-01)	&	2.81	&	6.11(-01)	&	3.03	&	5.45(-01)	&	2.70	&	9.45(-03)	\\
2.78	&	1.42(-01)	&	3.42	&	1.42(-01)	&	3.68	&	1.73(-01)	&	3.29	&	1.51(-02)	\\
3.38	&	9.31(-02)	&	4.16	&	9.21(-02)	&	4.48	&	1.23(-01)	&	4.00	&	3.73(-02)	\\
4.11	&	5.49(-02)	&	5.06	&	5.33(-02)	&	5.45	&	7.00(-02)	&	4.87	&	7.27(-02)	\\
5.00	&	3.46(-02)	&	6.15	&	3.28(-02)	&	6.63	&	3.35(-02)	&	5.92	&	1.22(-01)	\\
6.08	&	2.21(-02)	&	7.49	&	2.08(-02)	&	8.06	&	1.27(-02)	&	7.20	&	1.75(-01)	\\
7.40	&	1.42(-02)	&	9.11	&	1.34(-02)	&	9.81	&	5.12(-03)	&	8.76	&	2.08(-01)	\\
9.00	&	9.44(-03)	&	11.08	&	8.63(-03)	&	11.93	&	4.18(-03)	&	10.66	&	1.90(-01)	\\
10.95	&	6.42(-03)	&	13.48	&	5.52(-03)	&	14.52	&	5.42(-03)	&	12.96	&	1.18(-01)	\\
13.31	&	4.38(-03)	&	16.39	&	3.50(-03)	&	17.66	&	7.02(-03)	&	15.77	&	4.39(-02)	\\
16.20	&	3.00(-03)	&	19.94	&	2.49(-03)	&	21.48	&	7.91(-03)	&	19.18	&	8.56(-03)	\\
19.70	&	2.26(-03)	&	24.26	&	2.41(-03)	&	26.13	&	7.04(-03)	&	23.33	&	8.11(-04)	\\
23.97	&	2.10(-03)	&	29.51	&	2.85(-03)	&	31.79	&	4.43(-03)	&	28.38	&	3.76(-05)	\\
29.15	&	2.22(-03)	&	35.89	&	3.22(-03)	&	38.67	&	1.75(-03)	&	34.53	&	9.31(-07)	\\
35.46	&	2.16(-03)	&	43.66	&	2.98(-03)	&	47.03	&	3.88(-04)	&	42.00	&	1.38(-08)	\\
43.14	&	1.66(-03)	&	53.11	&	2.01(-03)	&	57.21	&	4.62(-05)	&	51.09	&	1.30(-10)	\\
52.47	&	8.78(-04)	&	64.61	&	8.71(-04)	&	69.60	&	3.01(-06)	&	62.15	&	8.22(-13)	\\
63.83	&	2.90(-04)	&	78.59	&	2.18(-04)	&	84.66	&	1.16(-07)	&	75.60	&	3.55(-15)	\\
77.65	&	5.54(-05)	&	95.60	&	2.95(-05)	&	102.99	&	2.85(-09)	&	91.96	&	1.07(-17)	\\
94.45	&	5.93(-06)	&	116.30	&	2.17(-06)	&	125.28	&	4.74(-11)	&	111.87	&	2.31(-20)	\\
114.89	&	3.67(-07)	&	141.47	&	9.24(-08)	&	152.39	&	5.50(-13)	&	136.08	&	3.57(-23)	\\
139.76	&	1.40(-08)	&	172.09	&	2.47(-09)	&	185.37	&	4.56(-15)	&	165.53	&	4.01(-26)	\\
170.01	&	3.48(-10)	&	209.33	&	4.38(-11)	&	225.50	&	2.76(-17)	&	201.36	&	3.30(-29)	\\
206.81	&	5.92(-12)	&	254.64	&	5.34(-13)	&	274.30	&	1.26(-19)	&	244.95	&	2.01(-32)	\\
251.57	&	7.04(-14)	&	309.75	&	4.60(-15)	&	333.67	&	4.46(-22)	&	297.96	&	9.09(-36)	\\
306.02	&	5.97(-16)	&	376.80	&	2.85(-17)	&	405.89	&	1.24(-24)	&	362.45	&	2.78(-39)	\\
372.26	&	3.68(-18)	&	458.35	&	1.31(-19)	&	493.75	&	2.77(-27)	&	440.90	&	0.00(00)	\\

 \hline

\end{tabular}
\end{table*}
%-------------------------------------------------

%-------------------------Table ------------------------
\begin{table*}
\caption{Grain size distribution function $f(a)$ normalised to $N_{tot}(a)$ as a function of size $a$ at day 2000 for forsterite, alumina, carbon and pure silicon, and the clumpy 19 \Ms\ ejecta model. }
\label{taba4}
\centering
\begin{tabular}{r c r c r c r c}
\hline \hline
$a$ (\AA) & $f(a)$ & $a$ (\AA) & $f(a)$ & $a$ (\AA) & $f(a)$ & $a$ (\AA) & $f(a)$ \\
\hline
\multicolumn{2}{c}{Forsterite} & \multicolumn{2}{c}{Alumina} & \multicolumn{2}{c}{Carbon} & \multicolumn{2}{c}{Pure-Si} \\
\hline
3.33	&	6.55(-01)	&	3.45	&	6.55(-01)	&	3.93	&	8.85(-02)	&	2.46	&	7.16(-01)	\\
4.05	&	1.41(-01)	&	4.19	&	1.41(-01)	&	4.78	&	6.04(-06)	&	2.99	&	1.40(-01)	\\
4.93	&	8.77(-02)	&	5.10	&	8.82(-02)	&	5.81	&	7.10(-07)	&	3.63	&	7.86(-02)	\\
6.00	&	4.73(-02)	&	6.20	&	4.80(-02)	&	7.07	&	7.41(-11)	&	4.42	&	3.60(-02)	\\
7.30	&	2.69(-02)	&	7.55	&	2.74(-02)	&	8.60	&	2.14(-13)	&	5.38	&	1.66(-02)	\\
8.88	&	1.55(-02)	&	9.18	&	1.58(-02)	&	10.46	&	6.12(-14)	&	6.54	&	7.36(-03)	\\
10.80	&	9.00(-03)	&	11.17	&	9.33(-03)	&	12.72	&	4.28(-14)	&	7.96	&	3.15(-03)	\\
13.14	&	5.32(-03)	&	13.59	&	5.62(-03)	&	15.48	&	2.96(-14)	&	9.68	&	1.33(-03)	\\
15.98	&	3.22(-03)	&	16.53	&	3.46(-03)	&	18.83	&	1.30(-14)	&	11.77	&	5.67(-04)	\\
19.44	&	2.03(-03)	&	20.10	&	2.14(-03)	&	22.90	&	3.02(-15)	&	14.32	&	2.42(-04)	\\
23.64	&	1.35(-03)	&	24.46	&	1.27(-03)	&	27.86	&	1.54(-14)	&	17.42	&	9.81(-05)	\\
28.76	&	9.56(-04)	&	29.75	&	6.61(-04)	&	33.89	&	1.25(-11)	&	21.19	&	3.70(-05)	\\
34.99	&	7.31(-04)	&	36.19	&	2.83(-04)	&	41.23	&	2.72(-09)	&	25.78	&	1.35(-05)	\\
42.56	&	6.02(-04)	&	44.02	&	1.20(-04)	&	50.15	&	1.89(-07)	&	31.36	&	5.01(-06)	\\
51.77	&	5.26(-04)	&	53.55	&	8.15(-05)	&	61.00	&	5.26(-06)	&	38.14	&	1.84(-06)	\\
62.97	&	4.78(-04)	&	65.14	&	7.93(-05)	&	74.21	&	7.08(-05)	&	46.40	&	6.63(-07)	\\
76.60	&	4.41(-04)	&	79.24	&	7.83(-05)	&	90.27	&	5.37(-04)	&	56.44	&	2.38(-07)	\\
93.18	&	4.03(-04)	&	96.39	&	8.57(-05)	&	109.81	&	2.61(-03)	&	68.66	&	9.02(-08)	\\
113.35	&	3.44(-04)	&	117.25	&	1.22(-04)	&	133.57	&	8.91(-03)	&	83.52	&	3.85(-08)	\\
137.88	&	2.46(-04)	&	142.62	&	1.89(-04)	&	162.48	&	2.33(-02)	&	101.59	&	2.11(-08)	\\
167.73	&	1.33(-04)	&	173.49	&	2.69(-04)	&	197.65	&	4.93(-02)	&	123.58	&	1.81(-08)	\\
204.03	&	5.18(-05)	&	211.04	&	3.23(-04)	&	240.43	&	8.84(-02)	&	150.33	&	2.64(-08)	\\
248.19	&	1.83(-05)	&	256.72	&	3.04(-04)	&	292.47	&	1.37(-01)	&	182.87	&	5.71(-08)	\\
301.91	&	1.13(-05)	&	312.29	&	2.03(-04)	&	355.77	&	1.79(-01)	&	222.45	&	1.42(-07)	\\
367.26	&	1.18(-05)	&	379.88	&	8.49(-05)	&	432.77	&	1.88(-01)	&	270.60	&	3.38(-07)	\\
446.74	&	1.39(-05)	&	462.10	&	2.02(-05)	&	526.44	&	1.43(-01)	&	329.16	&	7.81(-07)	\\
543.44	&	1.62(-05)	&	562.11	&	2.83(-06)	&	640.39	&	6.97(-02)	&	400.41	&	1.74(-06)	\\
661.06	&	1.72(-05)	&	683.78	&	5.14(-07)	&	778.99	&	1.88(-02)	&	487.07	&	3.52(-06)	\\
804.14	&	1.51(-05)	&	831.78	&	2.58(-07)	&	947.60	&	2.55(-03)	&	592.49	&	6.18(-06)	\\
978.19	&	9.95(-06)	&	1011.80	&	1.18(-07)	&	1152.70	&	1.66(-04)	&	720.73	&	9.49(-06)	\\
1189.90	&	4.43(-06)	&	1230.80	&	3.21(-08)	&	1402.20	&	5.53(-06)	&	876.73	&	1.26(-05)	\\
1447.50	&	1.26(-06)	&	1497.20	&	4.80(-09)	&	1705.70	&	1.05(-07)	&	1066.50	&	1.40(-05)	\\
1760.70	&	2.44(-07)	&	1821.30	&	3.93(-10)	&	2074.90	&	1.23(-09)	&	1297.30	&	1.20(-05)	\\
2141.80	&	4.04(-08)	&	2215.40	&	1.86(-11)	&	2523.90	&	9.56(-12)	&	1578.10	&	7.12(-06)	\\
2605.40	&	9.76(-09)	&	2695.00	&	5.54(-13)	&	3070.20	&	5.03(-14)	&	1919.70	&	2.59(-06)	\\
3169.30	&	6.31(-09)	&	3278.30	&	1.10(-14)	&	3734.70	&	1.84(-16)	&	2335.20	&	5.23(-07)	\\
3855.30	&	5.74(-09)	&	3987.80	&	1.50(-16)	&	4543.10	&	4.75(-19)	&	2840.60	&	5.60(-08)	\\
4689.70	&	4.14(-09)	&	4850.90	&	1.45(-18)	&	5526.40	&	8.78(-22)	&	3455.40	&	3.28(-09)	\\
5704.80	&	1.99(-09)	&	5900.90	&	1.00(-20)	&	6722.50	&	1.18(-24)	&	4203.30	&	1.15(-10)	\\
6939.50	&	5.79(-10)	&	7178.00	&	5.08(-23)	&	8177.60	&	1.15(-27)	&	5113.10	&	2.61(-12)	\\
8441.50	&	9.74(-11)	&	8731.70	&	1.90(-25)	&	9947.50	&	8.33(-31)	&	6219.80	&	4.07(-14)	\\
10269.00	&	9.56(-12)	&	10622.00	&	5.31(-28)	&	12101.00	&	4.47(-34)	&	7566.00	&	4.50(-16)	\\
12491.00	&	5.71(-13)	&	12920.00	&	1.12(-30)	&	14720.00	&	1.78(-37)	&	9203.60	&	3.61(-18)	\\
15195.00	&	2.19(-14)	&	15717.00	&	1.80(-33)	&	17906.00	&	0.00(00)	&	11196.00	&	2.13(-20)	\\
18484.00	&	5.61(-16)	&	19119.00	&	2.25(-36)	&	21781.00	&	0.00(00)	&	13619.00	&	9.39(-23)	\\
22484.00	&	9.95(-18)	&	23257.00	&	2.20(-39)	&	26495.00	&	0.00(00)	&	16566.00	&	3.12(-25)	\\
%27351.00	&	1.25(-19)	&	28291.00	&	1.71(-42)	&	32230.00	&	0.00(00)	&	20152.00	&	7.87(-28)	\\
%33270.00	&	1.13(-21)	&	34414.00	&	0.00(00)	&	39206.00	&	0.00(00)	&	24514.00	&	1.52(-30)	\\
%40471.00	&	7.46(-24)	&	41862.00	&	0.00(00)	&	47692.00	&	0.00(00)	&	29820.00	&	2.26(-33)	\\

\hline

\end{tabular}
\end{table*}
%-------------------------------------------------

\begin{table*}
\caption{Grain size distribution function $f(a)$ normalised to $N_{tot}(a)$ as a function of size $a$ at day 2000 for pure-Mg, pure-Fe, iron sulphide and silicon carbide and the clumpy 19 \Ms\ ejecta model.}
\label{taba5}
\centering
\begin{tabular}{r c r c r c r c}
\hline \hline
$a$ (\AA) & $f(a)$ & $a$ (\AA) & $f(a)$ & $a$ (\AA) & $f(a)$ & $a$ (\AA) & $f(a)$ \\
\hline
\multicolumn{2}{c}{Pure-Mg} & \multicolumn{2}{c}{Pure-Fe} & \multicolumn{2}{c}{FeS} & \multicolumn{2}{c}{SiC} \\
\hline
2.28	&	6.37(-01)	&	2.81	&	6.99(-01)	&	3.03	&	6.56(-01)	&	2.70	&	2.00(-03)	\\
2.78	&	1.41(-01)	&	3.42	&	1.38(-01)	&	3.68	&	1.40(-01)	&	3.29	&	6.88(-07)	\\
3.38	&	8.95(-02)	&	4.16	&	7.97(-02)	&	4.48	&	8.63(-02)	&	4.00	&	2.31(-06)	\\
4.11	&	5.00(-02)	&	5.06	&	3.88(-02)	&	5.45	&	4.61(-02)	&	4.87	&	3.60(-05)	\\
5.00	&	2.99(-02)	&	6.15	&	2.01(-02)	&	6.63	&	2.61(-02)	&	5.92	&	3.23(-04)	\\
6.08	&	1.84(-02)	&	7.49	&	1.09(-02)	&	8.06	&	1.50(-02)	&	7.20	&	1.79(-03)	\\
7.40	&	1.15(-02)	&	9.11	&	6.16(-03)	&	9.81	&	8.67(-03)	&	8.76	&	6.78(-03)	\\
9.00	&	7.35(-03)	&	11.08	&	3.43(-03)	&	11.93	&	5.06(-03)	&	10.66	&	1.92(-02)	\\
10.95	&	4.75(-03)	&	13.48	&	1.81(-03)	&	14.52	&	3.00(-03)	&	12.96	&	4.33(-02)	\\
13.31	&	3.12(-03)	&	16.39	&	9.07(-04)	&	17.66	&	1.85(-03)	&	15.77	&	8.17(-02)	\\
16.20	&	2.08(-03)	&	19.94	&	4.49(-04)	&	21.48	&	1.25(-03)	&	19.18	&	1.32(-01)	\\
19.70	&	1.39(-03)	&	24.26	&	2.22(-04)	&	26.13	&	1.04(-03)	&	23.33	&	1.84(-01)	\\
23.97	&	9.42(-04)	&	29.51	&	1.07(-04)	&	31.79	&	1.19(-03)	&	28.38	&	2.09(-01)	\\
29.15	&	6.46(-04)	&	35.89	&	4.92(-05)	&	38.67	&	1.60(-03)	&	34.53	&	1.79(-01)	\\
35.46	&	4.48(-04)	&	43.66	&	2.12(-05)	&	47.03	&	2.01(-03)	&	42.00	&	1.02(-01)	\\
43.14	&	3.12(-04)	&	53.11	&	8.75(-06)	&	57.21	&	2.09(-03)	&	51.09	&	3.35(-02)	\\
52.47	&	2.15(-04)	&	64.61	&	3.56(-06)	&	69.60	&	1.63(-03)	&	62.15	&	5.68(-03)	\\
63.83	&	1.43(-04)	&	78.59	&	1.46(-06)	&	84.66	&	8.50(-04)	&	75.60	&	4.65(-04)	\\
77.65	&	9.02(-05)	&	95.60	&	6.24(-07)	&	102.99	&	2.65(-04)	&	91.96	&	1.89(-05)	\\
94.45	&	5.78(-05)	&	116.30	&	3.06(-07)	&	125.28	&	4.84(-05)	&	111.87	&	4.20(-07)	\\
114.89	&	4.81(-05)	&	141.47	&	2.10(-07)	&	152.39	&	1.14(-05)	&	136.08	&	5.66(-09)	\\
139.76	&	5.74(-05)	&	172.09	&	2.34(-07)	&	185.37	&	1.36(-05)	&	165.53	&	4.95(-11)	\\
170.01	&	7.58(-05)	&	209.33	&	3.43(-07)	&	225.50	&	2.15(-05)	&	201.36	&	2.90(-13)	\\
206.81	&	9.11(-05)	&	254.64	&	6.16(-07)	&	274.30	&	3.02(-05)	&	244.95	&	1.17(-15)	\\
251.57	&	9.20(-05)	&	309.75	&	1.37(-06)	&	333.67	&	3.57(-05)	&	297.96	&	3.32(-18)	\\
306.02	&	7.19(-05)	&	376.80	&	3.07(-06)	&	405.89	&	3.36(-05)	&	362.45	&	6.73(-21)	\\
372.26	&	3.99(-05)	&	458.35	&	6.05(-06)	&	493.75	&	2.26(-05)	&	440.90	&	9.82(-24)	\\
452.83	&	1.53(-05)	&	557.56	&	1.03(-05)	&	600.61	&	9.68(-06)	&	536.33	&	1.05(-26)	\\
550.84	&	5.11(-06)	&	678.24	&	1.52(-05)	&	730.61	&	2.40(-06)	&	652.42	&	8.18(-30)	\\
670.06	&	2.07(-06)	&	825.04	&	1.91(-05)	&	888.74	&	3.28(-07)	&	793.62	&	4.74(-33)	\\
815.09	&	7.70(-07)	&	1003.60	&	1.97(-05)	&	1081.10	&	2.50(-08)	&	965.40	&	2.05(-36)	\\
991.51	&	1.79(-07)	&	1220.80	&	1.51(-05)	&	1315.10	&	1.13(-09)	&	1174.30	&	0.00(00)	\\
1206.10	&	2.26(-08)	&	1485.10	&	7.66(-06)	&	1599.70	&	3.22(-11)	&	1428.50	&	0.00(00)	\\
1467.20	&	1.54(-09)	&	1806.50	&	2.30(-06)	&	1946.00	&	6.12(-13)	&	1737.70	&	0.00(00)	\\
1784.70	&	6.00(-11)	&	2197.50	&	3.76(-07)	&	2367.20	&	8.10(-15)	&	2113.80	&	0.00(00)	\\
2171.00	&	1.46(-12)	&	2673.10	&	3.27(-08)	&	2879.50	&	7.73(-17)	&	2571.30	&	0.00(00)	\\
2640.90	&	2.39(-14)	&	3251.70	&	1.60(-09)	&	3502.80	&	5.51(-19)	&	3127.90	&	0.00(00)	\\
3212.50	&	2.71(-16)	&	3955.50	&	4.78(-11)	&	4260.90	&	3.01(-21)	&	3804.90	&	0.00(00)	\\
3907.80	&	2.18(-18)	&	4811.60	&	9.45(-13)	&	5183.10	&	1.30(-23)	&	4628.40	&	0.00(00)	\\
4753.60	&	1.28(-20)	&	5853.00	&	1.29(-14)	&	6305.00	&	4.45(-26)	&	5630.20	&	0.00(00)	\\
5782.50	&	5.48(-23)	&	7119.90	&	1.25(-16)	&	7669.60	&	1.21(-28)	&	6848.80	&	0.00(00)	\\
7034.10	&	1.75(-25)	&	8660.90	&	8.81(-19)	&	9329.70	&	2.59(-31)	&	8331.10	&	0.00(00)	\\
8556.50	&	4.20(-28)	&	10535.00	&	4.56(-21)	&	11349.00	&	4.30(-34)	&	10134.00	&	0.00(00)	\\

\hline

\end{tabular}
\end{table*}
%-------------------------------------------------

\end{document}